\documentclass[twocolumn,superscriptaddress,showpacs,floatfix,aps,prb]{revtex4}

\usepackage{graphicx}
\usepackage{bm}
\usepackage{amsmath}
\usepackage{dcolumn}%

\newcommand{\mb}{\ensuremath{\mu_{\text{B}}}}

\newcolumntype{/}{D{/}{/}{2,2}}  
\newcolumntype{.}{D{.}{.}{0}}  

\begin{document}

\title{Magnetic anisotropic effects and electronic correlations in
  MnBi ferromagnet}

\author{V.P. Antropov}

\affiliation{ Ames Laboratory USDOE, Ames, IA 50011}

\author{V.N. Antonov}

\affiliation{ Ames Laboratory USDOE, Ames, IA 50011}

\affiliation{Institute of Metal Physics, 36 Vernadsky Street, 03142
Kiev, Ukraine}

\author{L.V. Bekenov}

\affiliation{Institute of Metal Physics, 36 Vernadsky Street, 03142
Kiev, Ukraine}

\author{A. Kutepov}

\affiliation{ Ames Laboratory USDOE, Ames, IA 50011}

\author{G. Kotliar}

\affiliation{Department of Physics and Astronomy, Rutgers University,
Piscataway, New Jersey 08854, USA}

\date{\today}

\begin{abstract}

The electronic structure and numerous magnetic properties of MnBi
magnetic systems are investigated using local spin density
approximation (LSDA) with on-cite Coulomb correlations (LSDA+$U$)
included. We show that the inclusion of Coulomb correlations provides
a much better description of equilibrium magnetic moments on Mn atom
as well as the magnetic anisotropy energy (MAE) behavior with
temperature and magneto-optical effects. We found that the inversion
of the anisotropic pairwise exchange interaction between Bi atoms is
responsible for the observed spin reorientation transition at 90
K. This interaction appears as a result of strong spin orbit coupling
on Bi atoms, large magnetic moments on Mn atoms, significant $p-d$
hybridization between Mn and Bi atoms, and it depends strongly on
lattice constants. A better agreement with the magneto-optical Kerr
measurements at higher energies is obtained. We also present the
detailed investigation of the Fermi surface, the de Haas-van Alphen
(dHvA) effect and the X-ray magnetic circular dichroism in MnBi.

\end{abstract}

\pacs{75.50.Cc, 71.20.Lp, 71.15.Rf}

\maketitle



\section{Introduction.}

MnBi is an intriguing ferromagnetic material, both magnetically and
structurally. Manganese alloys usually tend to exhibit
antiferromagnetic order, because they have nearly half-filled 3$d$
bands, but MnBi is one of the few known ferromagnetic manganese
compounds which can be used as a permanent magnet. \cite{YKY+01} It is
other interesting magnetic properties include an extraordinarily large
Kerr rotation, \cite{DIT+92} with a Curie temperature above room
temperature (RT), \cite{Heikes55} a large perpendicular anisotropy in
thin films at RT, \cite{RuGu00} and a high coercivity that increases
with temperature. \cite{YYJC+02} The low-temperature phase (LTP) of
MnBi is ferromagnetic and has the hexagonal NiAs structure. With
increasing temperature, the material remains ferromagnetic up to 628 K
and then undergoes a coupled structural and magnetic phase transition
to a paramagnetic high-temperature phase (HTP). The HTP is a
disordered NiAs phase where 10-15\% of the large bipyramidal
interstitial sites are occupied by Mn atoms. \cite{ChSt74} Rapid
cooling of HTP MnBi yields a quenched high-temperature phase, which is
also ferromagnetic with even larger uniaxial magneto-crystalline
anisotropy energy (MAE), but smaller magnetization and Curie
temperature.

At RT MnBi is known to have an extremely high MAE (K$\sim$107
ergs/cm$^3$). This decreases rapidly, however, rapidly with decreasing
temperature and vanishes at $\sim$90 K ($T_{SR}$). \cite{AHF+67} The
experiments indicated the presence of a spin-reorientation transition
during this temperature decrease. \cite{YYJC+02} Among known hard
magnetic materials, MnBi is one of few alloys where the coercive field
increases with increasing temperature, reflecting the magnetic
anisotropy trend.

Historically, the ferromagnetic nature of manganese-bismuth alloys was
first reported by Heusler around 1904. \cite{Heusler04} In 1914 Bekier
considered the formation of a phase MnBi as probable; the phase
results from a peritectic reaction at 450 $^{\circ}$C between pure
manganese and the melted alloy containing 9 \% of manganese.
\cite{Bek14} Parravano and Perretl \cite{PaPe15} established the phase
diagram for this system and isolated crystals containing 19.9 \%
manganese, which they considered to be the MnBi phase. Hilpert and
Dieckmann \cite{HiHa11} noted the strong ferromagnetism of these
alloys and placed the Curie temperature at around 360-380
$^{\circ}$C. Furst and Halla later concluded from X-ray studies that a
single compound was present with the structure Mn$_2$Bi. \cite{FuHa38}
Montignie, however, showed that MnBi represented the only stable
compound. \cite{Mon38} In further studies by Halla and Montignie, the
same results were obtained. \cite{HaMO39}

The most comprehensive studies of this material were performed by
Guillaud in 1943 in Strasbourg. As a part of his PhD thesis, he was the
first to prepare the hexagonal MnBi compound and study its numerous
magnetic properties. \cite{book:Guill43} In addition to measuring the
saturation moment, MAE, and Curie temperature, he established the
dependence of the high coercive force of MnBi on its magnetic
anisotropy and reduced particle size. He also was first to observe the
spin reorientation and a corresponding increase of magnetic anisotropy
with temperature. Because of these considerations, around 60 years
ago, MnBi was chosen for the investigation by the US Naval Ordnance
Laboratory. As a result, a new permanent magnetic alloy "Bismanol" was
developed. \cite{Adams53} Bismanol has very high coercive force and
moderate energy density, making it a good choice for small electric
motors. However, due to oxidation and corrosion problems, Bismanol has
not been used much as a practical magnet.

Nevertheless, studies of both fundamental and applied properties
relevant for permanent magnetism have never been abandoned and this
material attracted attention of new generation of researchers. Hihara
T and Y. K\"oi \cite{HiKo70} studied the temperature dependence of the
easy axis of the magnetization in MnBi using the nuclear magnetic
resonance method. They found that for high temperatures above 142 K,
the easy axis of magnetization is along the $c$ direction. As the
temperature is decreased between the $T_{SR1}$=142 K and $T_{SR}$=90 K
(spin canting interval), the polar angle $\theta$ is gradually
deviates from the $c$ axis to $\theta_{exper} \sim 37^{\circ}$. The
magnetization flops into the $ab$ basal plane at 90 K. This spin
reorientation was also observed by neutron diffraction
\cite{AHF+67,YYJC+02} and magnetization measurements on MnBi single
crystals. \cite{SCS74}

Nearly simultaneously with Bismanol's development, MnBi became the
subject of another research activity after being recognized that, in
the form of thin films, it had quite favorable properties with regard
to potential applications for MO recording
\cite{WSF+57,May58,May60,Chen71} (see references prior 1988 and
discussion in Ref. \onlinecite{book:Bus88}). The MO properties of MnBi
were measured by several authors. \cite{DIT+92,HZL+93,DiUc96,HMW+98}
The most extensive study was carried out by Di {\it et al.}
\cite{DIT+92,DiUc96} Their measured Kerr angle spectrum for MnBi has
peaks at 1.84 eV and 3.35 eV. The former have a relatively large
magnitude of 2.31$^{\circ}$ at 85 K.

The electronic band structure of MnBi has been calculated by several
authors.
\cite{CoGr85,WaHu93,SaJa96,OAK+96a,UKH96,KoKu96,KoKu97,BOS+99,RDJ+99,TTB00,YYJC+02,MSZ+04,ZhDa04,WKD+06,LAK+07,SaEr08,KaFe10,GAC+11,KTL+11}
The optical and MO spectra of MnBi have been calculated in Refs.
\onlinecite{OAK+96a,SaJa96,KoKu96,KoKu97,RDJ+99}. However, there are
still disagreements about the interpretation of the MO spectra. The
major disagreement concerns the theoretical description of a high
energy peak of the Kerr rotation observed experimentally at 3.35 eV.

K\"ohler and K\"ubler \cite{KoKu96,KoKu97} obtained only one peak at
1.8 eV. They hypothesized that the thin-film samples may had
considerable impurities from materials in contact with them. They
found that oxygen, as an impurity, produced a second peak, but
energies of both peaks were not in good agreement with the
experiment. Ravindran {\it et al.}, \cite{RDJ+99} on the other hand,
did find a second peak in their calculated Kerr-angle spectrum for
pure MnBi. Oppeneer {\it et al.}  \cite{OAK+96a} obtained a large
negative peak at 1.8 eV. This is in agreement with experiment. They
then found only a shoulder around 3.4 eV. Here, the experimental data
has a pronounced peak. Since the data of Di {\it et al.}
\cite{DIT+92,DiUc96} were taken on a sample with the composition
Mn$_{1.22}$Bi, Oppeneer {\it et al.}  \cite{OAK+96a} simulated this
material and found a calculated Kerr-angle spectrum with a similar but
weaker peak at 1.8 eV and a second peak at 4.3 eV. This is higher in
energy than that for MnBi. Previous studies of MO properties of MnBi
were performed in the LDSA approximation and value of magnetic moment
was significantly underestimated.

Below we provide theoretical explanation for the long-standing
experimental puzzles in the measured MO properties, coercivity and
spin orientation. We show that all the physical properties under
consideration can be properly described only taking into account SO
interaction and Coulomb electron-electron correlations.

The paper is organized as follows. The crystal structure of MnBi and
computational details are presented in Sec. II. Sec. III presents
results and discussions of electronic and magnetic structures,
magnetic moments, Fermi surface (FS), orbital dependence of the
cyclotron masses, extremal cross sections of the FS, MO Kerr spectra,
x-ray magnetic circular dichroism, and temperature dependence of the
MAE of MnBi. The results are summarized in Sec. IV.

\section{Crystal structure and calculation details.}

\subsection{Crystal structure.}

\begin{figure}[tbp!]
\begin{center}
\includegraphics[width=0.99\columnwidth]{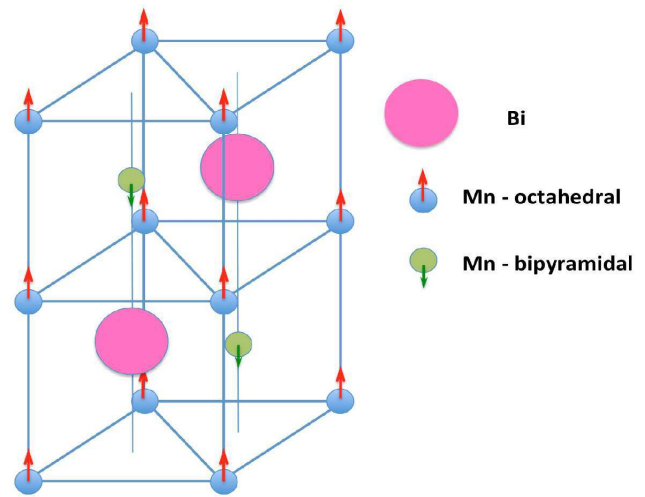}
\end{center}
\caption{\label{struc}(Color online) Crystal lattice of MnBi LTP
  phase. }
\end{figure}

Fig. \ref{struc} depicts the crystal and magnetic structure of
MnBi. It exhibits large trigonal-bipyramidal interstitial sites, which
may be occupied by dopant or Mn atoms. It is speculated that
octahedral Mn atoms are ferromagnetically coupled with the spin
parallel to the $c$-axis, and the bipyramidal Mn atoms are
antiferromagnetically coupled to the octahedral Mn atoms and result in
reduced net magnetization. \cite{AHF+67,BSK+98}

The LTP phase of MnBi is ferromagnetic and has a hexagonal NiAs-type
structure with unit cell dimensions $a$=4.29 \AA, $c$=6.13 \AA. On
heating above 360 $^{\circ}$C ($T_t$) there is a first order phase
transition to a paramagnetic \cite{Heikes55} state (HTP). The MnBi
compounds in the LTP phase have the NiAs structure ($P6\bar3/mmc$
symmetry, group number 194). The unit cell consists of two Mn atoms at
the Wyckoff 2$a$ sites (0, 0, 0) and (0, 0, $\frac{1}{2}$) and two Bi
atoms at the Wyckoff 2$c$ sites ($\frac{1}{3}$, $\frac{2}{3}$,
$\frac{1}{4}$) and ($\frac{2}{3}$, $\frac{1}{3}$, $\frac{3}{4}$).

\subsection{Calculation details}
\label{sec:Calc}

\paragraph{Magneto-optical properties and x-ray magnetic circular dichroism.}

For the polar Kerr magnetization geometry and a crystal of tetragonal
symmetry, where both the fourfold axis and the magnetization $\rm{ \bf
  M}$ are perpendicular to the sample surface and the $z$-axis is
chosen to be parallel to them, the dielectric tensor is composed of
the diagonal $\varepsilon_{xx}$ and $\varepsilon_{zz}$, and the
off-diagonal $\varepsilon_{xy}$ components in the form

\begin{equation}
\mbox{\boldmath$\varepsilon$}=\left(
\begin{array}{ccc}
\varepsilon_{xx} & \mbox{$\varepsilon_{xy}$} & 0 \\
\mbox{$-\varepsilon_{xy}$} & \mbox{$\varepsilon_{xx}$} & 0 \\
0 & 0 & \mbox{$\varepsilon_{zz}$}
\end{array}
\right) .
\label{defeps_pol}
\end{equation}

The various elements $\hat{\varepsilon}_{\alpha\beta}$ are composed of
real and imaginary parts as follows:
$\hat{\varepsilon}_{\alpha\beta}=\varepsilon_{\alpha\beta}^{(1)}+ i
\varepsilon_{\alpha\beta}^{(2)}$, where $\alpha ,\beta \equiv x,y,z$,
$\varepsilon_{xx}=(n + i k)^{2}$, and $n$ and $k$ are refractive index
and extinction coefficient, respectively.  The optical conductivity
tensor $\hat{\sigma}_{\alpha\beta}=\sigma_{\alpha\beta}^{(1)}+ i
\sigma_{\alpha\beta}^{(2)}$ is related to the dielectric tensor
$\varepsilon _{\alpha \beta }$ through the equation

\begin{equation}
 \hat{\varepsilon}_{\alpha
  \beta} (\omega ) = \delta_{\alpha \beta} +\frac{4 \pi
  i}{\omega}\hat{\sigma}_{\alpha\beta} (\omega ).
\label{defsigeps}
\end{equation}

The Kerr rotation $\theta$ and ellipticity $\eta$ are expressed as
follow: \cite{book:Sch92}

\begin{equation}
\theta +i\eta \approx \frac{-\varepsilon_{xy}}
{(\varepsilon _{xx}-1)\sqrt{\varepsilon _{xx}}} %
\,.
\label{defkerr}
\end{equation}

The optical conductivity of MnBi has been computed from the energy
band structure by means of the Kubo-Greenwood \cite{Kub57}
linear-response expression: \cite{WC74}

\begin{eqnarray}
\sigma _{\alpha \beta }(\omega ) &=&\frac{-ie^{2}}{m^{2}\hbar
 V_{uc}}\times
\nonumber \\
&&\sum_{{\bf k}}\sum_{nn^{\prime }}\frac{f(\epsilon
 _{n{\bf k}})-f(\epsilon
_{n^{\prime }{\bf k}})}{\omega _{nn^{\prime }}({\bf k})}\frac{\Pi
_{n^{\prime }n}^{\alpha }({\bf k})\Pi
 _{nn^{\prime }}^{\beta }({\bf k})}{%
\omega -\omega _{nn^{\prime }}({\bf k})+i\gamma }\,,
\label{defsig}
\end{eqnarray}
where $f(\epsilon _{n{\bf k}})$ is the Fermi function, $\hbar \omega
_{nn^{\prime }}({\bf k})\equiv \epsilon _{n{\bf k}}-\epsilon
_{n^{\prime } {\bf k}}$ is the energy difference of the Kohn-Sham
energies, $\epsilon _{n{\bf k}}$, and $\gamma $ is the lifetime
parameter, it is included to describe the finite lifetime of the
excited Bloch electron states. The $\Pi _{nn^{\prime }}^{\alpha }$ are
the dipole optical transition matrix elements. In a fully relativistic
description, these are given by

\begin{equation}
\mbox{\boldmath$\Pi$}_{nn^{\prime }}({\bf k})=\langle
 \psi _{n{\bf k}}|c%
\mbox{\boldmath$\alpha $}|\psi _{n^{\prime }{\bf k}}\rangle \,
\label{Pi}
\end{equation}
with the four-component Bloch electron wave function $\psi _{n{\bf
    k}}$, velocity of light $c$, and Dirac operator
\mbox{\boldmath$\alpha $}. The combined correction terms were also
taken into account in the optical matrix element calculations. A
detailed description of the optical matrix elements in the Dirac
representation is given in Refs.\ \onlinecite{ABP+93} and
\onlinecite{book:AHY04}.

Within the one-particle approximation, the absorption coefficient
$\mu^{\lambda}_j (\omega)$ for incident x-ray of polarization
$\lambda$ and photon energy $\hbar \omega$ can be determined as the
probability of electronic transitions from initial core states with
the total angular momentum $j$ to final unoccupied Bloch states

\begin{eqnarray}
\mu^j_{\lambda} (\omega) &=& \sum_{m_j} \sum_{n \bf k} | \langle
\Psi_{n \bf k} | \Pi _{\lambda} | \Psi_{jm_j} \rangle |^2 \delta (E
_{n \bf k} - E_{jm_j} - \hbar \omega ) \nonumber \\ &&\times \theta (E
_{n \bf k} - E_{F} ) \, ,
\label{mu}
\end{eqnarray}
where $\Psi _{jm_j}$ and $E _{jm_j}$ are the wave function and the
energy of a core state with the projection of the total angular
momentum $m_j$; $\Psi_{n\bf k}$ and $E _{n \bf k}$ are the wave
function and the energy of a valence state in the $n$-th band with the
wave vector $\bf k$; $E_{F}$ is the Fermi energy.  $\Pi _{\lambda}= -e
\mbox{\boldmath$\alpha $} \bf {a_{\lambda}}$ is the electron-photon
interaction operator in the dipole approximation (\ref{Pi}), $\bf
{a_{\lambda}}$ is the $\lambda$ polarization unit vector of the photon
vector potential, with $a_{\pm} = 1/\sqrt{2} (1, \pm i, 0),
a_{\parallel}=(0,0,1)$. Here, $+$ and $-$ denotes, respectively, left
and right circular photon polarizations with respect to the
magnetization direction in the solid. X-ray magnetic circular and
linear dichroism are given then by ($\mu_{+}-\mu_{-}$) and
($\mu_{\parallel}-(\mu_{+}+\mu_{-})/2$), respectively.

Usually, the exchange splitting of a core shell is small compared to
the band width of final valence states and can be neglected. However,
the exchange splitting of the $2p_{1/2,3/2}$ states of 3$d$ transition
metals may be as large as 0.4 eV. Then, transitions from core
states with different $m_j$ in Eq.~(\ref{mu}) occur at different
photon frequencies. This may lead to the appearance of giant XMLD in
cubic 3$d$ metals and it's strong dependence on the magnetization
direction. \cite{KuOp03}

At the core level, XMCD is not only element-specific but also orbital
specific. For 3$d$ transition metals, the electronic states can be
probed by the $K$, $L_{2,3}$ and $M_{2,3}$ X-ray absorption and
emission spectra. In Bi, one can use the $K$, $L_{2,3}$, $M_{2,3}$,
$M_{4,5}$, $N_{2,3}$, $N_{4,5}$, $N_{6,7}$, and $O_{2,3}$ spectra. For
unpolarized absorption spectra $\mu^{0}(\omega)$ allows only
transitions with $\Delta l= \pm 1, \Delta j= 0, \pm 1$ (dipole
selection rules). Therefore only electronic states with an appropriate
symmetry contribute to the absorption and emission spectra under
consideration.

\paragraph{Magnetocrystalline anisotropy.}

The internal energy of ferromagnetic materials depends on the
direction of spontaneous magnetization. Here we consider one part of
this energy, the MAE, which possesses the crystal symmetry of the
material. For the material exhibiting uniaxial anisotropy, such as a
hexagonal crystal, the MAE can be expressed as \cite{book:SW59}

\begin{eqnarray}
K = K_1 \sin^2 \theta + K_2 sin^4 \theta + K_3^{'} \sin^6 \theta
\nonumber \\ + K_3 \sin^2 \theta \cos[6(\phi + \psi)] + ... 
\end{eqnarray}
where $K_i$ is the anisotropy constant of the $i$th order, $\theta$
and $\phi$ are the polar angles of the Cartesian coordinate system
where the $c$ axis coincides with the $z$ axis (the Cartesian
coordinate system was chosen such that the $x$ axis is rotated through
90$^{\circ}$ from the a hexagonal axis) and $\psi$ is the phase angle.

Here, we study MAE caused only by the SO interaction and define it as
the difference between two self-consistently calculated and fully
relativistic total energies for two different magnetic field
directions, $K = E(\theta)$ - $E(<0001>)$.

\paragraph{Calculation details}
\label{sec:Calc_details}

The calculations presented in this work were performed using the
spin-polarized fully relativistic LMTO method \cite{APS+95} (denoted
further as LSDA+SO). To understand the influence of the SO interaction
on the MO properties and MAE, we used the scalar relativistic magnetic
Hamiltonian with SO coupling added variationally. \cite{And75} The
basis consisted of $s$, $p$, $d$, and $f$ LMTO's. The {\bf k-}space
integration was performed with an improved tetrahedron
method. \cite{BJA94} To attain good convergence in total energy, a
large number of {\bf k} points has to be used in the calculations. To
resolve the difference in total energies and to investigate the
convergence, we used 12008 and 18986 {\bf k} points in the irreducible
part of the Brillouin zone. This corresponds to 46656 and 74088 {\bf
  k} points in full zone.

\paragraph{Treatment of the Coulomb correlations}
\label{sec:Calc_U}

It is well known that the LSDA fails to describe the electronic
structure and properties of the systems in which the interaction among
the electrons is strong. In recent years, more advanced methods of
electronic structure determination such as LSDA plus self-interaction
corrections, \cite{PZ81} the LSDA+$U$ \cite{AZA91} method, GW
approximation, \cite{Hed65} and dynamical mean-field theory
\cite{MV89,PJF95,GKK+96} have sought to remedy this problem and have
shown considerable success. Among them, the LSDA+$U$ method is the
simplest and most frequently used.  We used the "relativistic"
generalization of the rotationally invariant version of LSDA+$U$
method \cite{YAF03} which takes into account SO coupling so that the
occupation matrix of localized electrons becomes non-diagonal in spin
indexes.

The screened Coulomb $U$ and exchange $J$ integrals enter the LSDA+$U$
energy functional as external parameters and have to be determined
independently. We tried several approximations to obtain Hubbard $U$
in this work and decided on the value $U$=4 eV and $J$=0.97 eV. These
are used throughout the paper.

The value of $U$ can be estimated from the photo-emission spectroscopy
and x-ray Bremsstrahlung isochromat spectroscopy experiments. Because
of difficulties with unambiguous determination of $U$ it can be
considered as a parameter of the model. Its value can therefore be
adjusted to achieve the best agreement of the results of LSDA+$U$
calculations with photoemission or optical spectra. \cite{BABH00}
While the use of an adjustable parameter is generally considered an
anathema among first principles practitioners, the LSDA+$U$ approach
does offer a plausible and practical method to approximately treat
strongly correlated orbitals in solids. The Hubbard $U$ and exchange
parameter $J$ can be determined from supercell LSDA calculations using
Slater's transition state technique \cite{AG91,SDA94} or from
constrained LSDA calculations (cLSDA). \cite{DBZ+84,SDA94,PEE98}
Recent extensions of the cLSDA method may be found in Refs.
\onlinecite{CoGi05} and \onlinecite{NAY+06}. The cLSDA method,
however, is known from early on to yield values of $U$ that are too
large in some cases. \cite{AKJ+06} For example, Anisimov and
Gunnarsson \cite{AG91} computed the effective on site Coulomb
interaction in metallic Fe and Ce. For Ce the calculated Coulomb
interaction was found to be about 6 eV in good agreement with
empirical and experimental estimates ranging from 5 to 7 eV. The
result for Fe (also about 6 eV) was surprisingly high since $U$ was
expected to be in the range of 2-3 eV for elemental transition metals,
with the exception of Ni. \cite{AJS97,MSH84} We applied the cLSDA
method to MnBi and obtained $U$=4.57 eV, $J$=0.97 eV.

Another method for determining the effective interaction is a scheme
based on the random-phase approximation (RPA). Early attempts of this
can be found in Refs. \onlinecite{SpAr98} and \onlinecite{Kot00}. A
method for calculating the Hubbard $U$, called the constrained RPA
(cRPA) scheme was proposed by Aryasetiawan {\it et al.} \cite{AIG+04}
some years ago. Subsequently, a combined cLSDA and cRPA method was
also proposed. \cite{SoIm05} The main merit of the cRPA method over
currently available methods is that it allows for a precise
elimination of screening channels. They are instead to be included in
a more sophisticated treatment of the model Hamiltonian.\cite{MAI09}
This method allows easy access to obtaining not only on-site matrix
elements but also off-site matrix elements as well as
screened-exchange matrix elements. These are usually taken to be the
atomic value. Another merit is the possibility of obtaining the
frequency-dependent Hubbard $U$, and may prove to be important. The
cRPA method has now been applied to a number of systems with success.
\cite{AKJ+06,MiAr08,NAI08,MPV+08}

We have calculated an effective interaction for MnBi using a general
method of cRPA proposed by Aryasetiawan {\it et al.} \cite{AIG+04} In
this method one divides the full polarizability
$P(\mathbf{r},\mathbf{r}';\nu)$ into two parts: the first part
$P^{d}(\mathbf{r},\mathbf{r}';\nu)$ which is defined by all
transitions strictly between chosen (usually strongly correlated)
eigen states of one-particle Hamiltonian, and the second part
$P^{r}(\mathbf{r},\mathbf{r}';\nu)=P(\mathbf{r},\mathbf{r}';\nu)-P^{d}(\mathbf{r},\mathbf{r}';\nu)$.
After that, the effective interaction $U(\mathbf{q};\nu)$ can
equivalently be found either using $P^{d}(\mathbf{q};\nu)$ (we write
the equations in $\mathbf{q}$ space and omit product-basis indexes for
brevity here)

\begin{align}\label{U_f}
\big[1+W(\mathbf{q};\nu)P^{d}(\mathbf{q};\nu)\big]U(\mathbf{q};\nu)=W(\mathbf{q};\nu),
\end{align}
or using $P^{r}(\mathbf{q};\nu)$

\begin{align}\label{U_r}
\big[1-V(\mathbf{q})P^{r}(\mathbf{q};\nu)\big]U(\mathbf{q};\nu)=V(\mathbf{q}).
\end{align}

At this point, it is extremely important to understand that such
defined partial polarizabilities $P^{d}$ and $P^{r}$ both should
possess proper asymptotic behavior at small $\mathbf{q}$ (namely
their $P^{\mathbf{q}}_{\mathbf{G}=\mathbf{G}'=0}$ components in
plane wave representation should be proportional to $q^{2}$) in
order to cancel the corresponding $1/q^{2}$ divergency in
$V(\mathbf{q})$ or $W(\mathbf{q};\nu)$ as it is seen from the above
equations.

It is easy to show that the above requirement is automatically
satisfied when one uses band eigen states of some Hamiltonian of the
solid to construct the polarizabilities. But the orbital character of
the selected bands does not always corresponds perfectly to the
character of orbitals in which we are interested.

Therefore in our present work, we follow to the procedure: we pick up
the bands with proper orbital character (of course we admix some
amount of wrong orbital character in this way). Having the proper
bands ($d$-bands) picked up for every $\mathbf{k}$-point we calculate
$d$-polarizability $P^{d}(\mathbf{r},\mathbf{r}')$ which comes only
from these bands. We perform this in $\mathbf{q}$-space

\begin{align}\label{pol_f}
P^{d,\textbf{q}}_{ij}(\tau)=&-\sum_{\textbf{k}}\sum_{\lambda\lambda'\lambda''\lambda'''}
\sum_{\alpha}\nonumber\\&\times<M^{\textbf{q}}_{i}\Psi^{\alpha\textbf{k-q}}_{\lambda'''}|\Psi^{\alpha\textbf{k}}_{\lambda}>
G^{\alpha\mathbf{k}}_{\lambda\lambda'}(\tau)\nonumber\\& \times
<\Psi^{\alpha\textbf{k}}_{\lambda'}|\Psi^{\alpha\textbf{k-q}}_{\lambda''}M^{\textbf{q}}_{j}>G^{\alpha\textbf{k-q}}_{\lambda''\lambda'''}
(\beta-\tau),
\end{align}
$\alpha$ is the spin index, and the summations are performed over the
group of $d$-bands only. In (\ref{pol_f}),
$G^{\alpha\mathbf{k}}_{\lambda\lambda'}(\tau)$ is a full GW Green's
function which we express in the basis of LDA-bands ( indexes
$\lambda,\lambda',\lambda'',\lambda'''$). The above $d$-polarizability
defines an effective interaction among $3d$-electrons
$U_{d}(\mathbf{r},\mathbf{r}')$ which we again calculate in
$\mathbf{q}$-space:
$U^{-1}_{d}(\mathbf{q})=W^{-1}(\mathbf{q})+P^{d}(\mathbf{q})$, and in
product basis representation.

Finally, we calculate matrix elements of such found $U_{d}$ in a
basis of atomic orbitals

\begin{align}\label{u_wan}
\mathcal{U}_{LL';L''L'''}(\nu)&= \sum_{\alpha\alpha'}\int d \mathbf{r}
\int d \mathbf{r}'
\nonumber\\ &\times\varphi^{*}_{L}\mathbf{r})\varphi_{L'}(\mathbf{r})U^{d}_{\mathbf{rr}'}(\nu)
\varphi^{*}_{L'''}(\mathbf{r}')\varphi_{L''}(\mathbf{r}') ,
\end{align}
where the integrations are performed over the corresponding
muffin-tin sphere where $3d$-orbitals are defined. As atomic
orbitals we use the solutions of radial equations inside of the MT
spheres (which we define when we solve LDA equations).

Having calculated the full matrix we average it to get the effective
value $U$

\begin{align}\label{u_wan1}
U(\nu)=\frac{1}{N^{2}_{d}}\sum_{LL'}\mathcal{U}_{LL;L'L'}(\nu),
\end{align}
where $N_{d}$ is the degeneracy of $d$-set.

As it follows our method doesn't use mapping onto Wannier
representation like in the work by Miyake {\it et al.}  \cite{MAI89}
We think however for materials with well localized $d$-electrons, the
resulted effective interaction is not sensitive to such details. In
our opinion, much more important is the question - which Green's
function (LDA, QP, or self consistent GW) is used when one calculates
full- and $d$-polarizabilities.

\begin{figure}[tbp!]
\begin{center}
\includegraphics[width=0.99\columnwidth]{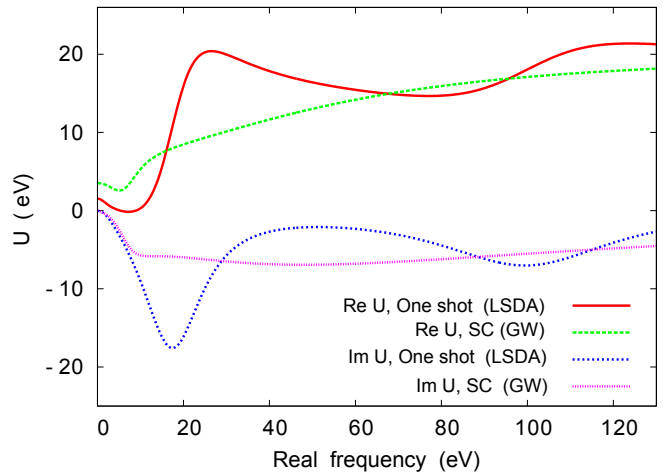}
\end{center}
\caption{\label{MnBi_U_rax}(Color online) Real and imaginary parts of
  the partially screened effective interaction $U$ for $3d$ shell of
  Mn for ferromagnetic MnBi analytically continued to the real
  axis. One-shot GW (starting with LSDA) and self-consistent GW
  results. }
\end{figure}

We have performed one shot (starting with LSDA) and fully
self-consistent GW calculations. We have used the mesh of k-points
$6\times6\times6$ in Brillouin zone (results obtained with
$4\times4\times4$ k-mesh differ very little). Green's function was
expanded over the FLAPW band Bloch states. The number of bands in this
expansion was 222-240 depending on $\mathbf{k}$-point. Inside the MT
spheres we have expanded the functions of the fermionic type (Green's
function and self energy) in spherical harmonics up to
$L_{max}=4$. Bosonic functions (polarizability and interaction) have
been expanded up to $L_{max}=6$. In the interstitial region, each
function was expanded in plane waves. We have used more plane waves
for the bosonic functions (~340) than for fermionic ones. Our full
basis size to expand the bosonic functions (add muffin-tin and
interstitial) was about 880 functions depending again on the point in
Brillouin zone. We calculated the effective $U$ as a function of
Matsubara frequency and then we analytically continued it to the real
frequency axis.

We present in Fig. \ref{MnBi_U_rax} our calculated effective
interaction $U$ for MnBi obtained in one shot GW and self-consistent
GW calculations. The one shot result for $U$ at zero frequency is
about 2 eV whereas the result from self-consistent calculation is
approximately 3.6 eV, both of them are smaller than the cLSDA value of
4.57 eV. As mentioned above, the cLSDA method usually overestimates
the values of $U$ for transition metals. On the other hand, the cRPA
underestimates Hubbard $U$. Therefore, in our calculations we use
$U$=4 eV and $J$=0.97 eV, and we use them throughout the paper.

\section{Energy band structure.}

\begin{figure}[tbp!]
\begin{center}
\includegraphics[width=0.99\columnwidth]{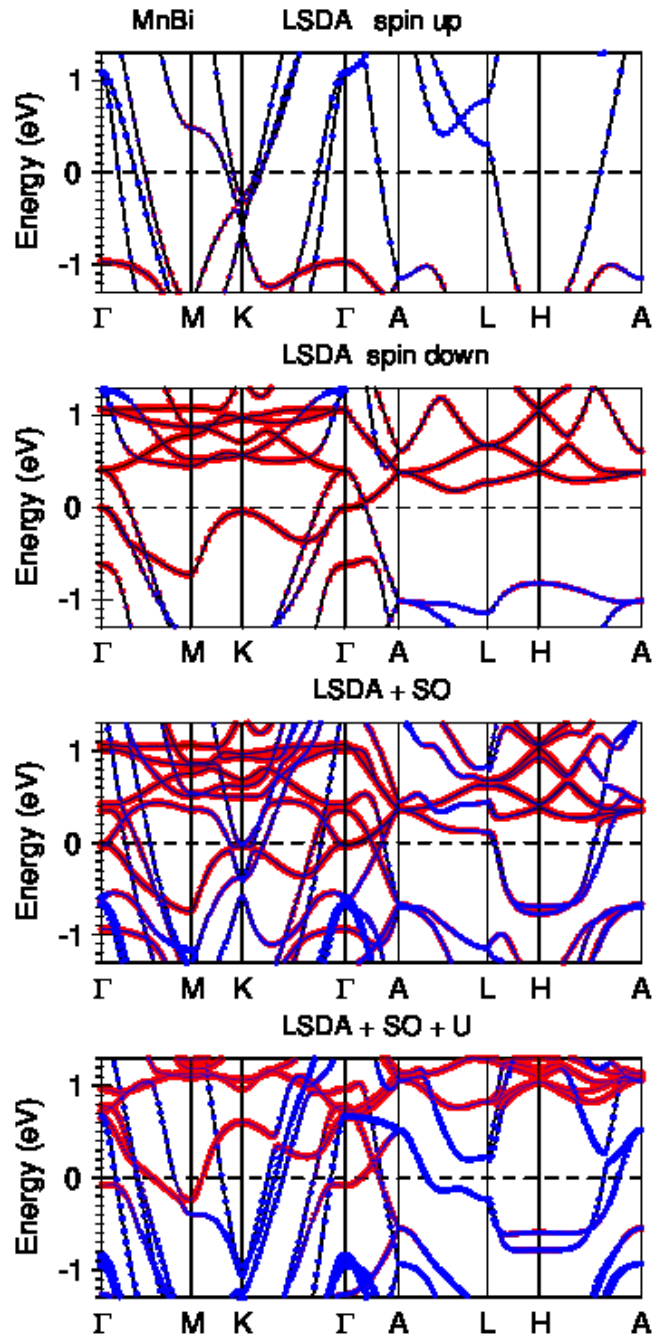}
\end{center}
\caption{\label{Ek}(Color online) Energy band structure of MnBi in
  close vicinity of the Fermi level using "fat band" representation: a
  non-relativistic (two upper panels); fully relativistic (third panel
  from the top) and fully relativistic LSDA+SO+$U$ (lower panel)
  energy bands. }
\end{figure}

Figure \ref{Ek} shows the spin-polarized energy band structure of MnBi
calculated in the LSDA without SO interaction (two upper panels),
fully relativistic Dirac approximation (LSDA+SO, third panel from the
top), and a fully relativistic Dirac LSDA+SO+$U$ approximation (lower
panel). The LSDA+SO results are in good agreement with previous LSDA
studies. \cite{LAK+07,SaEr08,KaFe10,GAC+11,KTL+11} In "fat band"
representation, the open red circles show the Mn 3$d$ character of the
wave function in each {\bf k} point. Closed blue circles indicate the
Bi 6$p$ character. The larger circle corresponds to the larger
contribution of the corresponding character in the wave function for a
given {\bf k} point.

The splitting of the energy bands in the $H$ and $A$ symmetry point in
the $-$0.6 eV to $-$1 eV energy interval is enhanced more than two
times after the inclusion of the Coulomb repulsion (compare third
panel from the top with lower panel in Fig. \ref{Ek}). Due to the
shift of Mn 3$d$ states from the Fermi level in the LSDA+SO+$U$
approach the character of the electronic states at the Fermi level are
changed towards the decreasing of a Mn 3$d$ partial contribution at
the Fermi level.

\begin{figure}[tbp!]
\begin{center}
\includegraphics[width=0.99\columnwidth]{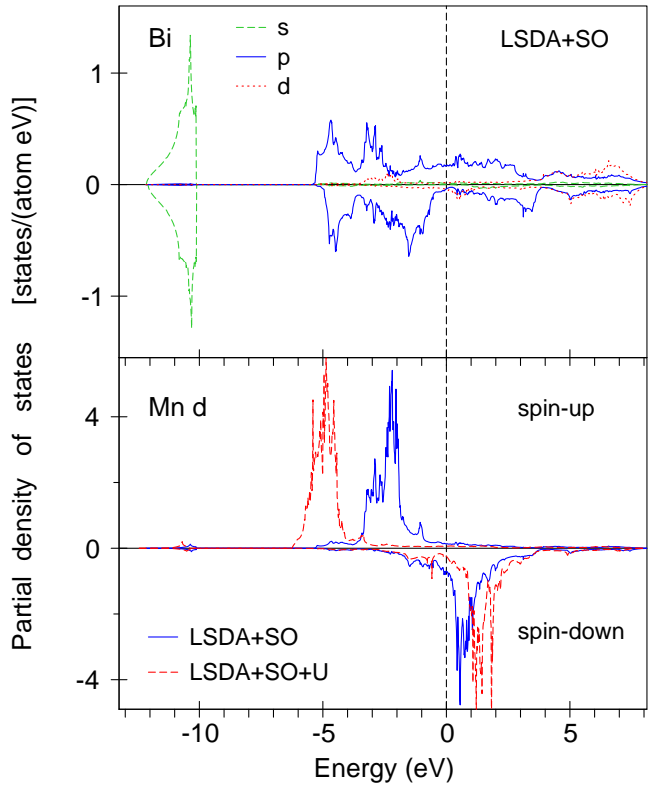}
\end{center}
\caption{\label{PDOS}(Color online) The LSDA+SO and LSDA+SO+$U$
  partial densities of states for MnBi. }
\end{figure}

Figure \ref{PDOS} shows partial densities of states for MnBi
calculated within LSDA+SO as well as LSDA+SO+$U$. The Mn $d$-states
are split by the on-site exchange interaction into nearly completely
filled majority-spin and unoccupied minority-spin states. The crystal
field at the Mn site ($D_{3d}$ point symmetry) causes the splitting of
$d$-orbitals into a singlet $a_{1g}$ ($3z^2-1$) and two doublets $e_g$
($yz$ and $xz$) and $e_{g1}$ ($xy$ and $x^2-y^2$). Bi 6$s$ states
situated at the $-$12.2 eV to $-$10.3 eV below the Fermi level. Bi
6$p$ states occupy the $-$5.2 eV to 7.5 eV energy range and strongly
hybridize with Mn 3$d$ states in the $-$4 eV to 3 eV energy range. The
spin splitting of Bi $p$-states is quite small. The LSDA+$U$ Mn 3$d$
partial DOSs are also presented in Fig. \ref{PDOS}. Usually the
failure of the LSDA method generally occurs toward the right end of
the 3$d$ transition-metal series. For Mn, which is in the middle of
3$d$ series, no strong correlation would be expected. As can be seen
below, however, the correlation effects are quite important in MnBi
for a correct description of the MO properties as well as the MAE.

Our fully relativistic LSDA band structure calculations produce $M_s$
of 3.572 $\mb$ at the Mn site in MnBi. The $M_s$ of $-$0.114 $\mb$
induced at the Bi site is antiparallel to that of Mn. The orbital
magnetic moment ($M_l$) at the Mn and Bi sites are equal to 0.156
$\mb$ and $-$0.028 $\mb$, respectively. An additional empty sphere
also carries small $M_s$ and $M_l$ of $-$0.015 $\mb$ and 0.001 $\mb$,
respectively. The net magnetic moment in the fully relativistic LSDA
band structure calculations is equal to 3.572 $\mb$. Experimental
numbers have been obtained for different samples and for samples of
different purity in a range from 3.82 $\mb$, \cite{ZHW+91} to 4.25
$\mb$. \cite{YYC+11} For most pure samples, the moment is close to 4.1
$\mb$ and compares favorably with our LSDA+SO+$U$ moment (4.172
$\mb$). The spin and orbital magnetic moments in the LSDA+SO+$U$
approach at the Mn site are equal to 4.224 $\mb$ and 0.125 $\mb$,
respectively, and $M_s$=$-$0.134 $\mb$ and $M_l$=$-$0.030 $\mb$ at the
Bi site.

The opposite sign of $M_s$ on Mn and Bi atoms can be understood
already from Fig. \ref{PDOS}. In the LSDA, the DOS of Mn $d$-states
near the Fermi level is much larger for a nearly empty spin down
electronic band. The latter band just starts to populate in Mn
reflecting typical more-than-half-filled $d$-band behavior. For
$p$-states of Bi (upper panel of Fig. \ref{PDOS}) the situation is
opposite: the population of $p$-states and their induced magnetic
moments are relatively small, with a larger DOS for $p$-states for
spin up.

The presence of a large amount of Mn spin down and Bi spin up
electrons at and near the Fermi level creates favorable conditions for
the appearance of the large transversal $p-d$ transitions induced by
SO coupling.

\section{Ground state properties.}

\subsection{Fermi surface}

In this section we present the topology of Fermi surface of MnBi as
well as the de Haas-van Alphen (dHvA) extremal cross-sections and
cyclotron masses.

\begin{figure}[tbp!]
\begin{center}
\includegraphics[width=0.7\columnwidth]{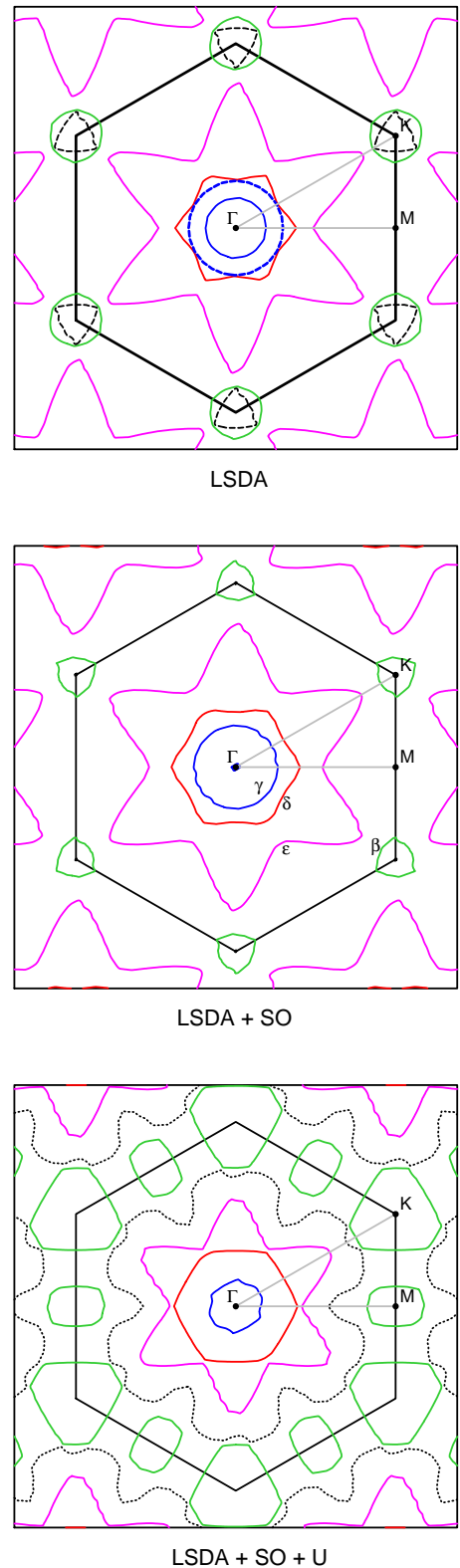}
\end{center}
\caption{\label{FS_cs}(Color online) The calculated cross sections of
  MnBi FS in the plane perpendicular to the $z$ direction $k_z$=0
  using a non-relativistic approach (upper panel), LSDA+SO (middle
  panel) and LSDA+SO+$U$ ($U$=4 eV) (lower panel). }
\end{figure}

\begin{figure}[tbp!]
\begin{center}
\includegraphics[width=0.5\textwidth]{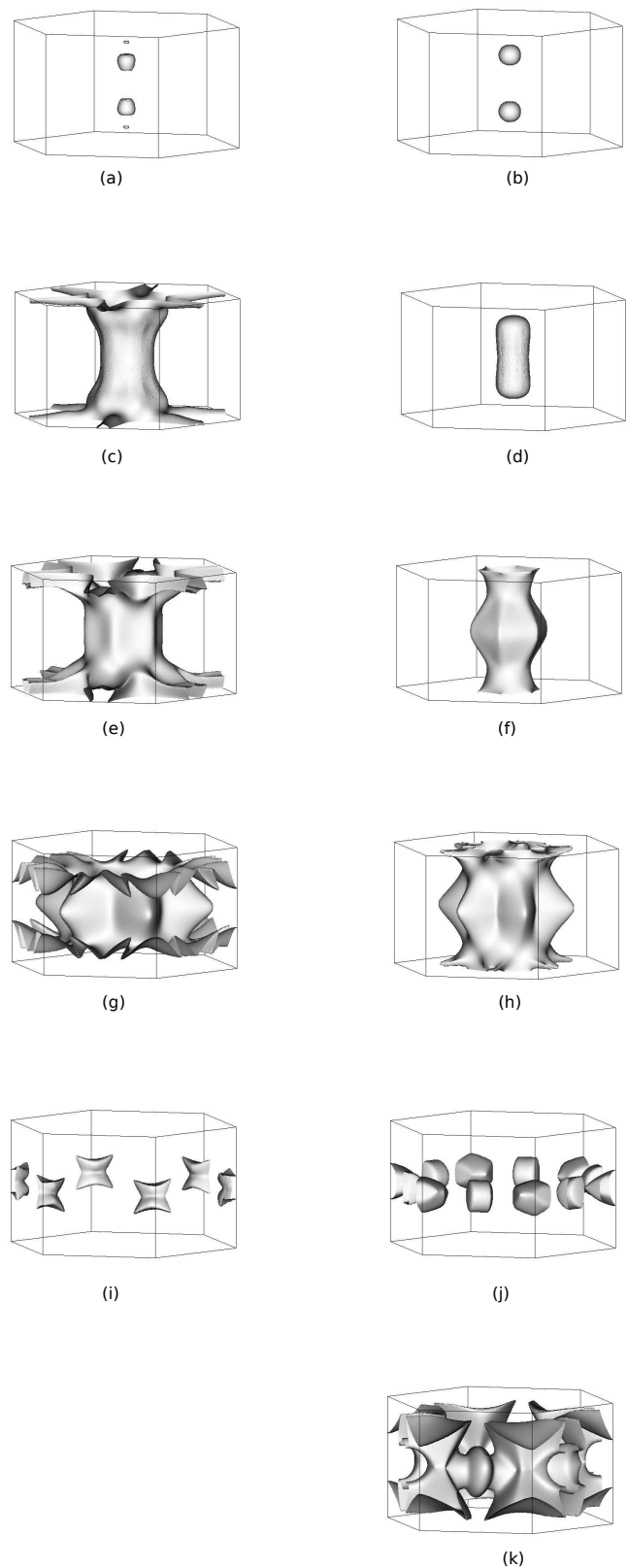}
\end{center}
\caption{\label{FS} (Color online) The LSDA+SO (left panels) and the
  LSDA+SO+$U$ (right panels) sheets of MnBi Fermi surface. }
\end{figure}

Figure \ref{FS_cs} shows the calculated cross sectional areas of MnBi
FS in the plane perpendicular to the $z$ direction $k_z$=0 crossed
$\Gamma$ symmetry point using a non-relativistic method (upper panel),
fully relativistic LSDA+SO (middle panel) and fully relativistic
LSDA+SO+$U$ (lower panel) approximations. Fig. \ref{FS} shows the
sheets of the FS in MnBi calculated with the LSDA+SO (left panels) and
LSDA+SO+$U$ (right panels) approximations. The inclusion of the SO
interaction changes the topology of the FS in MnBi
(Fig. \ref{FS_cs}). Instead of two sheets in the $K$ symmetry point in
the LSDA calculations we have only one electron FS. Besides, there are
four FS cross sections in the $k_z$=0 plane in the spin-polarized
calculations and three in the LSDA+SO calculations.

There are five sheets of the FS in the LSDA+SO calculations. Almost
spherical closed hole FS centered approximately at a half distance
between $\Gamma$ and $A$ symmetry points in Fig. \ref{FS}(a), has pure
Mn 3$d$ character. The 23, 24, and 25 hole FSs opened along the
$\Gamma-A$ direction with sixfold symmetry in Figs. \ref{FS}(c),
\ref{FS}(e), and \ref{FS}(g), respectively, are mostly due to Bi 6$p$
states with small amounts of Mn 3$d$ states mixed in. A closed
electron FS centered in the $K$ symmetry point in Fig. \ref{FS}(i) is
the mix of Mn 3$d$ and Bi 6$p$ character.

Inclusion of the Coulomb repulsion increases the size of the 22d hole
FS in Fig. \ref{FS}(b) and reconstruct the 24th and 25th hole FSs in
Fig. \ref{FS}(f)) and Figs. \ref{FS}(h)). It produces a new hole FS
sheet (see Fig. \ref{FS}(k)). Dashed black curve in the lower panel of
Fig. (\ref{FS_cs}), and an additional closed electron FSs centered in
$M$ symmetry point (lower panel of Fig. \ref{FS_cs}).

\begin{figure}[tbp!]
\begin{center}
\includegraphics[width=0.5\textwidth]{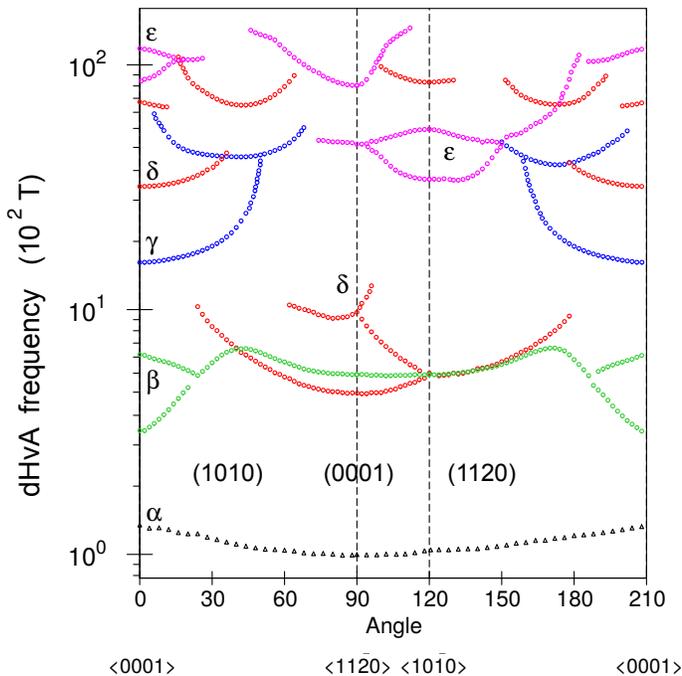}
\end{center}
\caption{\label{FS_dHvA_U} (Color online) The calculated angular
  dependence of the dHvA oscillation frequencies in MnBi using
  LSDA+SO+$U$ ($U$=4 eV) approximation.}
\end{figure}

\begin{figure}[tbp!]
\begin{center}
\includegraphics[width=0.7\columnwidth]{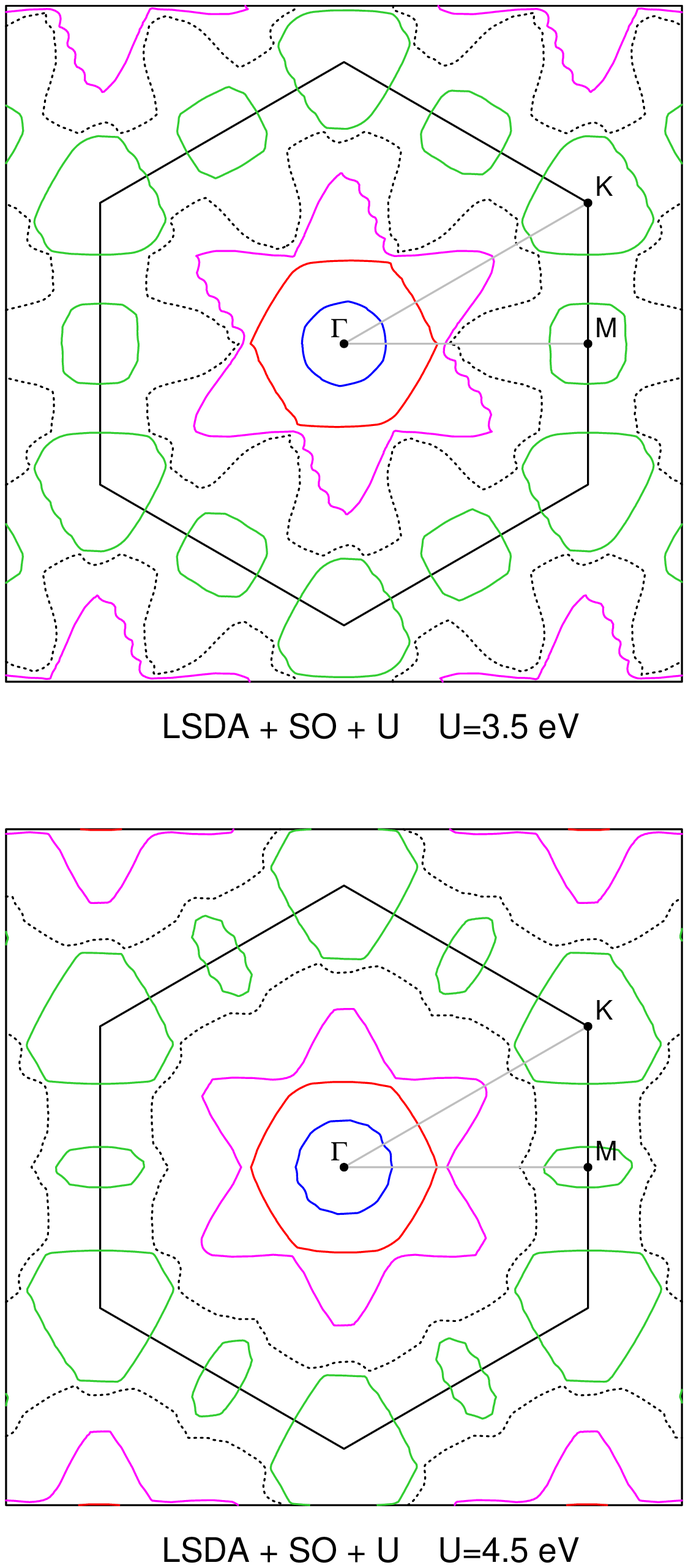}
\end{center}
\caption{\label{FS_cs_Uii}(Color online) The calculated cross sections
  of MnBi FS in the plane perpendicular to the $z$ direction $k_z$=0
  using fully relativistic LSDA+SO+$U$ method for $U$=3.5 eV (upper
  panel) and $U$=4.5 eV (lower panel). }
\end{figure}

Figure \ref{FS_dHvA_U} presents the angular variations of the
theoretically calculated dHvA frequencies in MnBi in the LSDA+SO+U
approximations for field direction in the ($10\bar10$), ($11\bar20$),
and (0001) planes. The obtained six different type orbits $\alpha$,
$\beta$, $\gamma$, $\delta$, $\varepsilon$, and $\sigma$ belong to the
FSs derived by the crossing of the 22nd, 23rd, 24th, 25th, 26th, and
27th energy bands, respectively. The $\alpha$ orbits situated at the
almost spherical closed hole FS which which centered at a half
distance between $\Gamma$ and $A$ symmetry points
(Figs. \ref{FS}(b)). Due to smallness and almost spherical shape of
these sheets the corresponding dHvA frequencies are rather small and
have almost constant angle dependence. The $\beta$ oscillations belong
to electron FS around the $K$ and $M$ points. These orbits split for
three separate $\beta_1$, $\beta_2$, and $\beta_3$ orbits. The
$\beta_3$ oscillations belong to the closed electron FS sheets around
the $M$ symmetry point. The $\beta_1$, and $\beta_2$ are at the
electron FS around the $K$ point (see Fig. \ref{FS_cs} lower
panel). The $\gamma$ and $\delta$ orbits exist in wide angle interval
at all the three planes. The highest dHvA frequencies were observed
for the $\varepsilon$ orbits situated at the hole surface derived from
the 26th energy band.

\begin{figure}[tbp!]
\begin{center}
\includegraphics[width=0.5\textwidth]{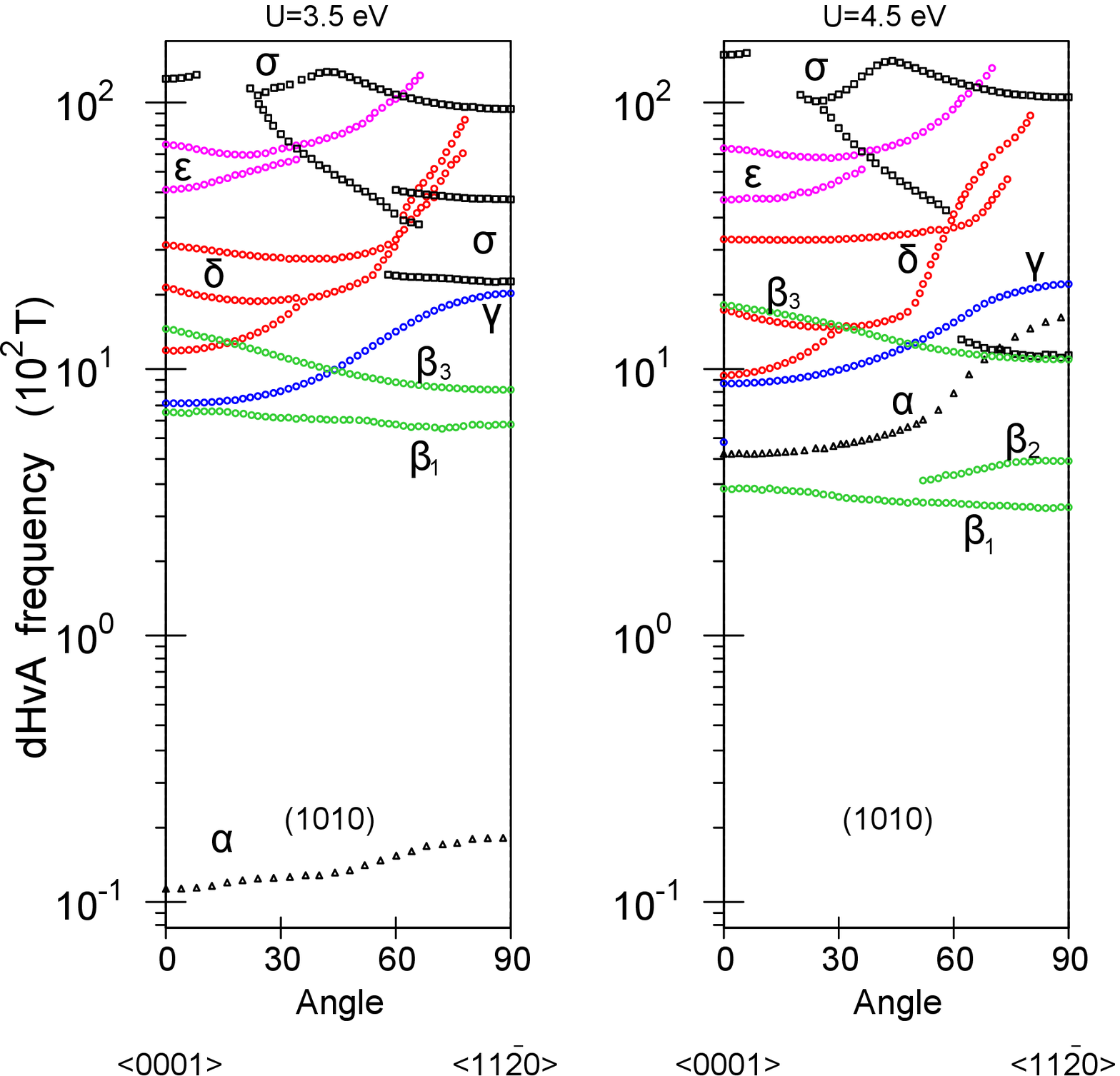}
\end{center}
\caption{\label{FS_dHvA_Uii} (Color online) The calculated angular
  dependence of the dHvA oscillation frequencies in MnBi using
  LSDA+SO+$U$ approximation for $U$=3.5 eV (left panel) and $U$=4.5 eV
  (right panel). }
\end{figure}

We also calculated the angular dependence of the cyclotron masses for
MnBi in the LSDA+SO+U approach (not shown). The masses for the
low-frequency oscillations $\alpha$ range from $-$1.0 $m_0$ to $-$0.65
$m_0$, and the dHvA $\beta$ orbits on the electron FS sheet around the
$K$ symmetry point possess relatively small cyclotron masses from 0.5
$m_0$ to 0.8 $m_0$. The $\delta$ orbits also have relatively small
cyclotron masses of $-$0.8 $m_0$ to $-$0.4 $m_0$. However, some
branches of the $\delta$ orbits possess cyclotron masses more than 2
$m_0$. The masses for the high-frequency oscillations $\varepsilon$
are large.

To show how sensitive are the calculations of the FS to the value of
Hubbard $U$, we run additional LSDA+SO+SO calculations with Hubbard
$U$=3.5 eV and 4.5 eV. Fig. \ref{FS_cs_Uii} shows the calculated cross
sections of MnBi FS for $U$=3.5 eV and $U$=4.5 eV. After comparing
this figure with Fig. \ref{FS_cs} for $U$=4.0 eV, we can conclude
qualitatively that all the three calculations produce similar FSs with
small changes in the size and the shape of some FS sheets.

However, the dHvA oscillations are quite sensitive to the value of
Hubbard $U$. The frequencies of the $\alpha$ orbits are significantly
reduced for $U$=3.5 eV and increased for $U$=4.5 eV in comparison with
$U$=4.0 eV calculations.  The $\delta$ and $\varepsilon$ orbits have
an opposite $U$ behavior, their frequencies are decreased with the
increase of $U$. On the other hand, the $\sigma$ and $\gamma$ orbits
are less sensitive to the value of $U$.

The experimental measurements of the dHvA effect is highly desired, it
will answer which value of Hubbard $U$ is realized in MnBi. From the
experimental point of view it would be no problem to measure the dHvA
oscillations in MnBi do due to relatively small cyclotron masses for
most dHvA orbits. However, a single crystal sample of a good quality
might be needed.

\subsection{Magneto-crystalline anisotropy.}

It has been established that such unique temperature dependence of the
coercivity and MAE in MnBi is determined by the thermal variation of
the lattice parameters $a$ and $c$. In the following section we
explain experimental observations by examining the dependence of
calculated total energy and MAE on the lattice geometry. We assume
that the finite temperature can be mimicked by the lattice constants
corresponding to this temperature. We confirm that spin reorientation
arises from a change of sign in MAE, which depends on the lattice
constants.

\begin{figure}[tbp!]
\begin{center}
\includegraphics[width=0.99\columnwidth]{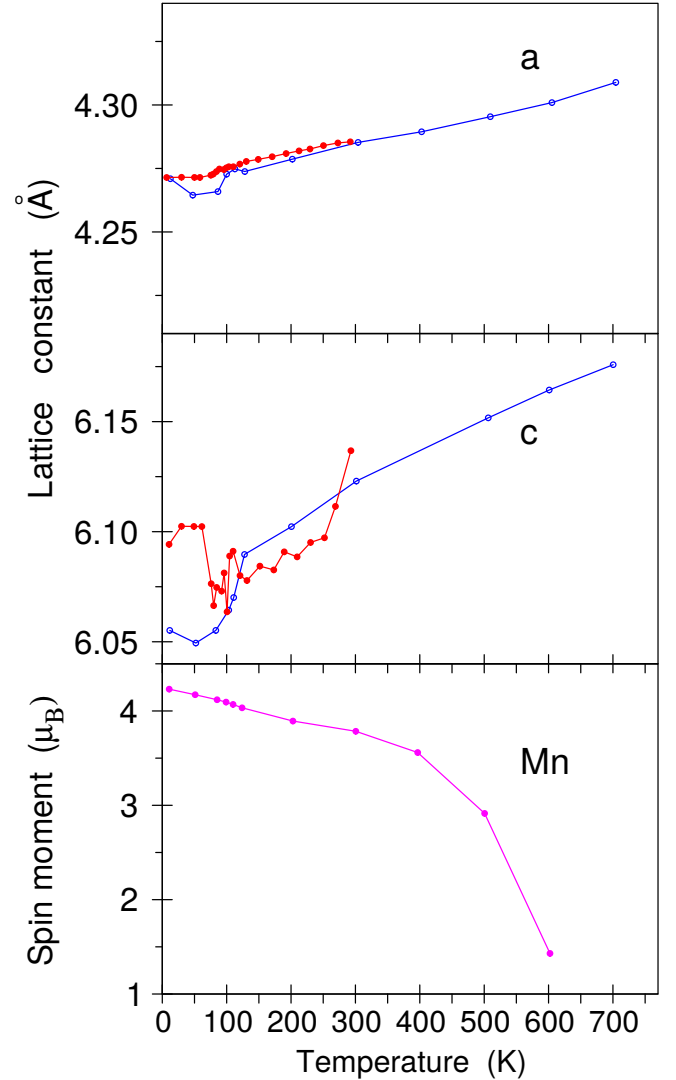}
\end{center}
\caption{\label{ac_M}(Color online) The temperature dependence of the
  lattice parameters $a$, and $c$ (upper and middle panels,
  respectively) of MnBi according to the Ref. \onlinecite{YYC+11}
  (blue curve) and Ref. \onlinecite{KMW08} (red curve). The lower
  panel shows temperature dependence of magnetization in MnBi.
  \cite{YYC+11} }
\end{figure}

Fig. \ref{ac_M} show the experimentally measured temperature
dependence of the lattice constants $a$ and $c$ according to the
Refs. \onlinecite{YYC+11} and \onlinecite{KMW08} together with the
magnetization \cite{YYC+11} in MnBi. Yang {\it et al.} \cite{YYC+11}
measured temperature dependence of the lattice parameters $a$, and $c$
in a wide temperature range from 10 to 700 K. Koyoma {\it et al.},
\cite{KMW08} on the other hand, used a smaller temperature interval
(10$-$300 K). They did use a very fine temperature mesh in the
vicinity of $T_{SR}$. Both measurements show similar behavior for the
lattice constant $a$, but strongly differ from each other in the
temperature behavior of lattice constant $c$. The results of Yang {\it
  et al.}  \cite{YYC+11} show a rather smooth decrease of parameter
$c$ with decreasing of $T$ below $T_{SR}$. Koyoma {\it et al.}
\cite{KMW08} found a discontinuous behavior of constant $c$ near
$T_{SR}$ (Fig. \ref{ac_M}).

\begin{figure}[tbp!]
\begin{center}
\includegraphics[width=0.99\columnwidth]{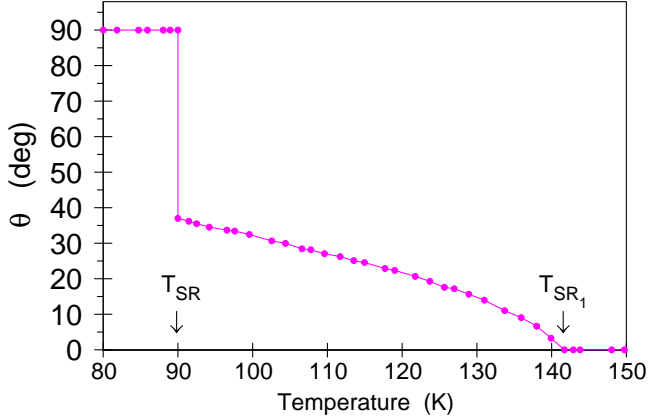}
\end{center}
\caption{\label{SR1}(Color online ) Temperature variation of the polar
  angle $\theta$ between the easy axis of the magnetization and the
  $c-$axis \cite{HiKo70} in MnBi. }
\end{figure}

\begin{figure}[tbp!]
\begin{center}
\includegraphics[width=0.99\columnwidth]{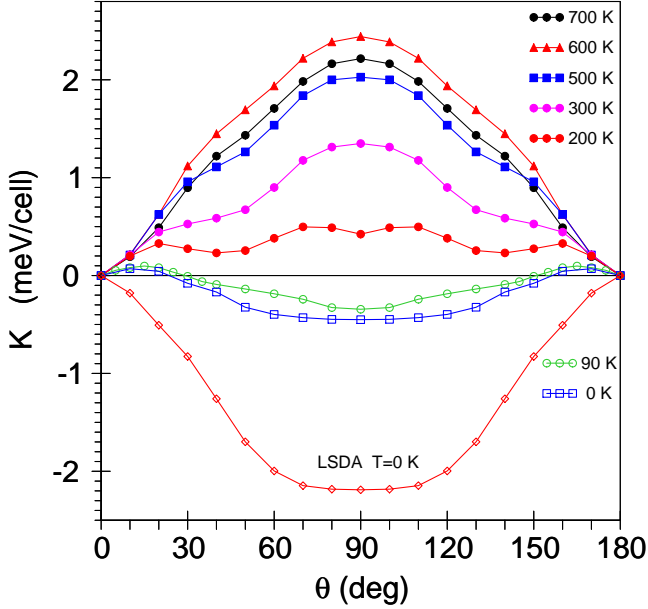}
\end{center}
\caption{\label{SW_T}(Color online) The MAE $K$ as a function of the
  polar angle $\theta$ and temperature in MnBi calculated in the
  LSDA+SO (open blue squares) and the LSDA+SO+$U$ methods. The
  temperature dependence of the lattice constants $a$ and $c$ is from
  Ref.  \onlinecite{YYC+11}. }
\end{figure}

Figure \ref {SW_T} shows the MAE as a function of the polar angle
$\theta$ and temperature calculated with the LSDA+SO and the
LSDA+SO+$U$ methods. Here we used the temperature dependence of the
lattice constants $a$ and $c$ obtained by Yang {\it et al.}
\onlinecite{YYC+11}. The LSDA+SO approach gives the value of MAE equal
to $-$2.2 meV/cell at zero temperature. This value is in good
agreement with a previous FPLAPW band structure calculation by
Ravindran {\it et al.} \cite{RDJ+99} ($-$2.0 meV/cell). However, both
of these values are an order of magnitude larger than the experimental
value of $-$0.13 meV/cell. \cite{SCS74,VSZ92} Besides, the LSDA+SO
approximation shows that the easy direction of the magnetization is in
the basal plane for any value of lattice constant $a$ and axial ratio
$c/a$ (meaning the entire temperature range) and, therefore, provides
no explanation of the spin-reorientation observed experimentally at
the $T_{SR} \sim$90 K. On the other hand, the LSDA+SO+$U$ approach
gives the value of MAE equal to $-$0.39 meV/cell at zero
temperature. This value is already in better agreement with the
experiment, but still nearly three times larger than the
experimentally estimated value of $-$0.13 meV/cell. \cite{SCS74,VSZ92}
Thus the inclusion of the Coulomb correlations provides a correct easy
magnetization direction along $c$ axis for the temperatures above
$T_{SR1}$ and in the plane below $T_{SR}$ for the experimental
parameters $a$ and $c$.

\begin{figure}[tbp!]
\begin{center}
\includegraphics[width=0.99\columnwidth]{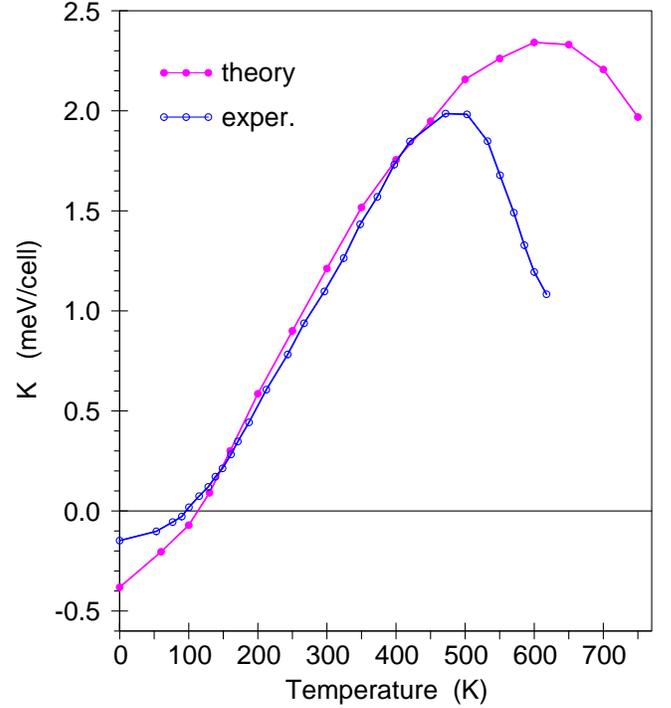}
\end{center}
\caption{\label{MAE_T}(Color online) Theoretically calculated
  temperature dependence of the MAE $K$ in MnBi using the LSDA+SO+$U$
  in comparison with the experiment. \cite{SCS74} }
\end{figure}

Figure \ref{MAE_T} presents the theoretically calculated temperature
dependence of the MAE in MnBi using the LSDA+SO+$U$ approximation in
comparison with the experiment. \cite{SCS74} The theoretical MAE is in
a very good agreement with the experiment in the 150 K to 450 K
temperature range. Thus, our calculations confirm the experimental
claim \cite{KMW08} that the unusual temperature dependence of MAE is
primarily due to a specific lattice thermal expansion.

To determine a major source of MAE we expanded MAE over SOC parameters
by direct varying these parameters on different sites and fitting the
resulting MAE with on-site and intersite contributions function. This
analysis has shown that the dominant contribution to the MAE variation
is produced by the anisotropic pairwise interaction between $p$-states
of Bi atoms. Thus, the physical reason for the spin orientation
observed in MnBi at 90K is the exchange striction, when anisotropic
Bi-Bi pair interaction changes its sign. Further studies of this
inversion of the anisotropic exchange in materials with spin
reorientation is needed.

With the temperature increase above RT, the experimentally measured
anisotropy energy increases and reaches its maximum at around 500 K,
and then rapidly decreases at higher temperatures. The LSDA+SO+$U$
results show the same temperature behavior. They, however, show higher
MAE in the maximum. Besides, the theoretically calculated maximum of
the MAE shifts towards higher temperatures (Fig. \ref{MAE_T}). Such
disagreement between theory and experiment might be due to the
magnetic spin disorder effect. The temperature dependence of
magnetization in MnBi measured by Yang {\it et al.}  \cite{YYC+11}
shows a drastic reduction of the magnetization from the 4.25 $\mb$ at
0 K to the 1.43 $\mb$ for 600 K. This presumably due to spin disorder
(see lower panel of Fig. \ref{ac_M}). Such an effect has not been
taken into account in our calculations.

\begin{figure}[tbp!]
\begin{center}
\includegraphics[width=0.99\columnwidth]{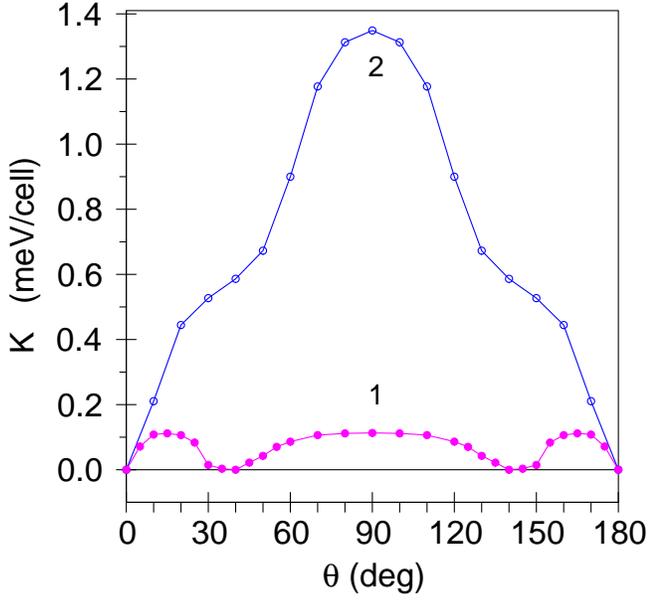}
\end{center}
\caption{\label{SW_a}(Color online) The LSDA+SO+$U$ calculations of
  the MAE $K$ as a function of the polar angle $\theta$ for the
  lattice constant $c$=6.123 \AA. This corresponds to the $T$=300 K
  and $a$=4.272 \AA\, (curve 1) and $a$=4.283 \AA\, (curve 2)
  corresponding to the $T$=100 K and 300 K,
  respectively. \cite{YYC+11} }
\end{figure}

We found a strong dependence of the MAE on in-plane lattice constant
$a$.  Fig. \ref{SW_a} shows the MAE as a function of the polar
angle $\theta$ for the lattice constant $c$=6.123 \AA. This
corresponds to $T$=300 K and $a$=4.272 \AA\, (curve 1) and
$a$=4.283 \AA\, (curve 2) corresponding to the $T$=100 K and 300 K,
respectively. \cite{YYC+11} Expansion of in-plane lattice constant $a$
by 0.01 \AA\, occurs from 100 K and 300 K increases the MAE by
approximately 1.2 meV. Corresponding results for $c$ parameter
expansion produce a much smaller result of 0.1 meV (Fig. \ref{SW_a}).

\begin{figure}[tbp!]
\begin{center}
\includegraphics[width=0.99\columnwidth]{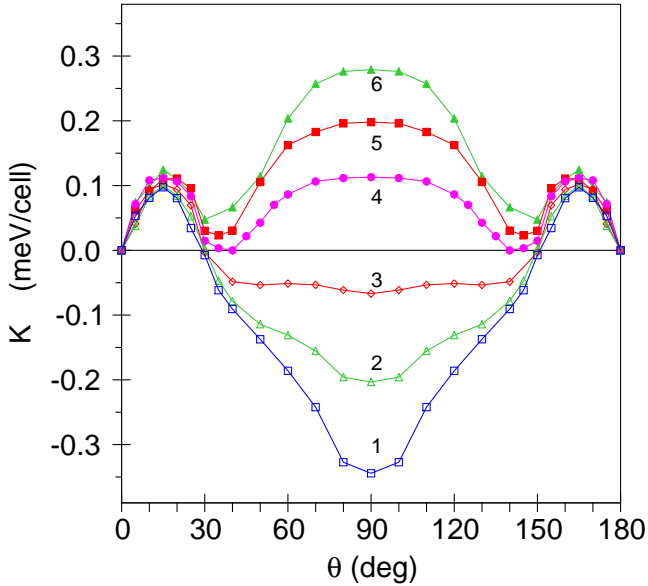}
\end{center}
\caption{\label{SW_c}(Color online) The LSDA+SO+$U$ calculations of
  the MAE $K$ as a function of the polar angle $\theta$ for the
  lattice constant $a$=4.274 \AA. This corresponds to the
  spin-reorientation ($T_{SR}$=90 K) \cite{KMW08} and $c$=6.09 \AA\,
  (curve 1), $c$=6.10 \AA\, (curve 2), $c$=6.11 \AA\, (curve 3),
  $c$=6.12 \AA\ (curve 4), $c$=6.13 \AA\, (curve 5), and $c$=6.14
  \AA\, (curve 6). }
\end{figure}

To investigate the MAE as a function of the polar angle $\theta$ in
the vicinity of the spin reorientation phase transition, we fixed the
lattice constant $a$ for the spin-reorientation temperature $a$=4.274
\AA\,\cite{KMW08} and vary $c$ from $c$=6.09 \AA\, to $c$=6.14 \AA\,
with a step of 0.01 \AA\, (Fig. \ref{SW_c}). For the lattice constants
$c$=6.09 \AA, $c$=6.10 \AA\,, and $c$=6.11 \AA\, (curves 1-3,
respectively), the easy magnetization direction is in the basal
plane. There are two local minima in the total energy for the $c$=6.12
\AA\, (curve 4): one along the $c$ direction and at the $\theta \sim$
41$^{\circ}$ with a barrier in between. The last angle is close to the
experimentally measured $\theta_{exper}=37 ^{\circ}$ at $T_{SR}$=90 K
where the magnetization flops into the $ab$ basal plane. \cite{HiKo70}
(see Fig. \ref{SR1}). For larger values of the lattice constant $c$,
the easy magnetization direction is along the $c$ direction in
agreement with experimental observation. We would like to point out
that the results presented in Fig. \ref{SW_c} have to be considered
only as qualitative ones because by fixing the constant $a$ and
varying $c$, the $c/a$ has altered the overall volume per unit cell.

The results shown in Fig. \ref{SW_c} lead to some interesting
conclusions. The angular dependence of the total energy demonstrates a
presence of a "double-well" potential. This fact leads to a hysteresis
phenomenon as a function of temperature. For instance, one can expect
a non-smooth dependence of the magnetization direction change with a
"sudden" switch of easy direction at different temperatures depending
on whether a cooling or heating process is being used. This
qualitatively explains the non analytical dependence of easy axes
observed in Ref. \onlinecite{HiKo70} (see Fig. \ref{SR1}). The total
energy shows a highly non-trivial angular dependence with several
minima. This leads to a hysteresis behavior of magnetization as a
function of temperature. This in turn creates a condition for a
non-continuous spin reorientation transition, that can be considered
as a planar to the uniaxial anisotropy phase transition.

We can conclude that the increase of MAE with temperature presented in
Fig. \ref{MAE_T} is mostly due to changing the in-plane lattice
constant $a$.

\begin{figure}[tbp!]
\begin{center}
\includegraphics[width=0.99\columnwidth]{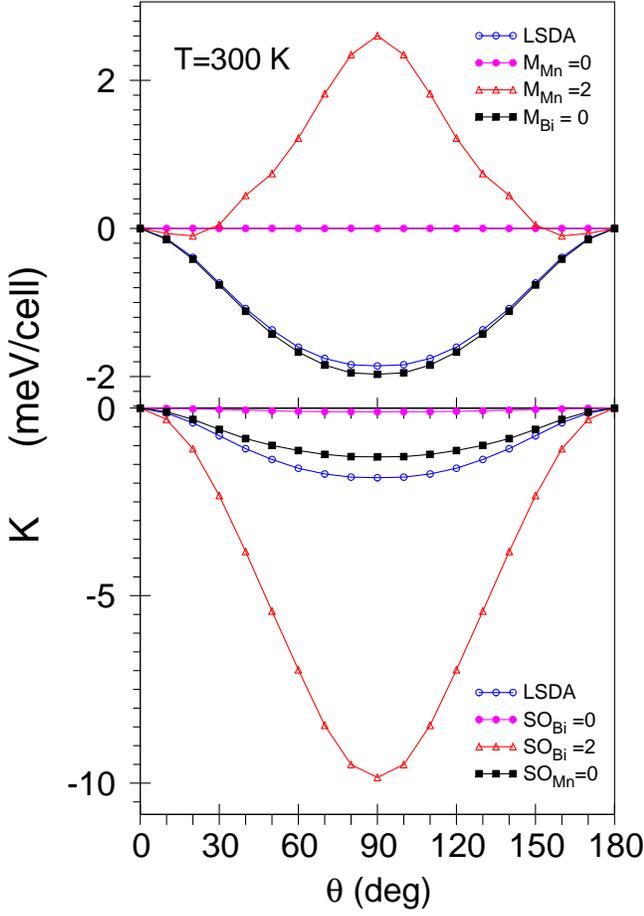}
\end{center}
\caption{\label{SW_H_SO}(Color online) The effect of scaling of the
  exchange splitting (upper panel) and value of SO constant (lower
  panel) on either Mn or Bi atoms on the MAE $K$ in MnBi(see the
  text). }
\end{figure}

We examine the dependence of the MAE on the exchange splitting and the
SO interaction. The exchange splitting and the SO coupling are studied
by scaling the corresponding terms in the Hamiltonian artificially
with a constant prefactor. This scaling can be atom dependent, i.e.,
within each atomic sphere. The outcomes of such constraining
calculations for the MAE in MnBi are shown in Fig. \ref{SW_H_SO}. In
the upper panel, the importance of the exchange splitting is
illustrated. When the exchange splitting on Bi is set to zero, the MAE
is barely modified. But when we performed such operation on Mn atom,
the MAE totally vanishes. Furthermore, an enhancement of the exchange
splitting on Mn by a factor of 2.0 (red open triangles) leads to a
correct easy magnetization along the $z$ direction.

The lower panel of Fig. \ref{SW_H_SO} shows the dependence on the SO
coupling. If we set SO coupling on Mn to zero, the MAE does not change
significantly (full black squares). On the other hand, when the SO
coupling on Bi is zero, the MAE almost disappears (magenta full
circles). The scaling of the SO coupling of Bi by a factor of 2.0
leads to an increase of the MAE by a factor of 5 confirming a dominant
contribution of the SO interaction at the Bi site to the large value
of MAE in this compound.

\begin{figure}[tbp!]
\begin{center}
\includegraphics[width=0.99\columnwidth]{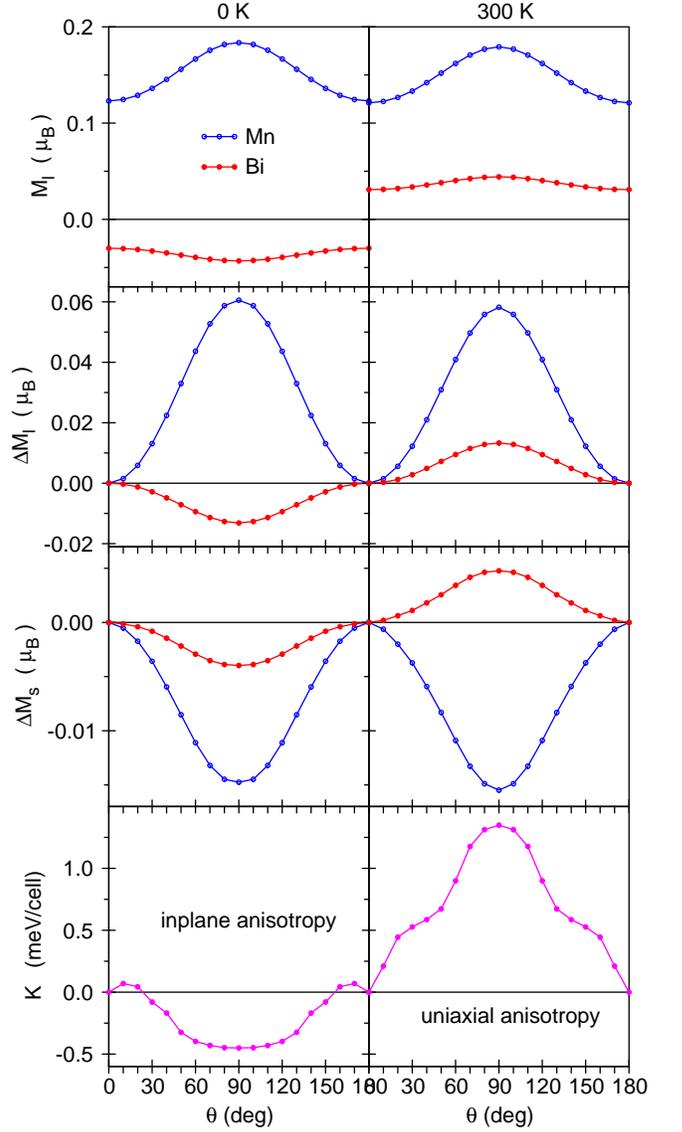}
\end{center}
\caption{\label{K_Ml_Ms}(Color online) The MAE, $M_l$, OMA ($\Delta
  M_l$) and SMA ($\Delta M_s$) for $T$= 0 K (left panel)and $T$= 300 K
  (right panel). }
\end{figure}

It is customary to relate the MAE with the anisotropy of $M_l$ (OMA).
\cite{YOC65,CEM94,AKA14} Fig. \ref{K_Ml_Ms} presents the MAE, $M_l$,
OMA, and the anisotropy of $M_s$ (SMA) for $T$= 0 K and $T$= 300
K. The $M_l$, OMA and SMA are larger at the Mn site than at the Bi
one. The OMA is four times larger than the SMA for both sites. The Bi
$M_l$ changes its sign through spin-reorientation transition,
therefore the inversion of the MAE through the spin-reorientation
transition is directly correlated with the inversion of $M_l$ at the
Bi site.

The MAE is approximately proportional to the OMA through expression $K
\sim \frac{1}{4} \lambda \Delta M_l$, \cite{YOC65,CEM94,AKA14,KBS13}
where $\lambda$ is the SO parameter ($\sim$0.041 eV for Mn and
$\sim$0.85 eV for Bi \cite{book:MCP+06}). Therefore, the major
contribution to the MAE is due to the OMA at the Bi site with some
contribution from the SMA. A direct proportionality of $K$ and OMA
above is directly related to the third Hund's rule for the more than
half-filled band. However, we argued in Ref. \onlinecite{AKA14} that a
difference of DOS at $E_f$ for the different spin channels can be
crucial for the sign of this proportionality. As we discussed above,
there are more spin up $p$-states of Bi atom at $E_f$. This in turn
leads to inverse proportionality between $K$ and OMA. Thus, a minimum
of orbital moment on Bi site corresponds to minimum of the total
energy, so $K \sim -\frac{1}{4} \lambda \Delta M_l$, that corresponds
to a third Hund's rule for a less than half filled band. This is
similar to the situation in CoPt and FePt, discussed in
Ref. \onlinecite{AKA14}.

Recently, Zarkevich {\it et al.} \cite{cm:ZWJ13} calculated the total
energy and MAE versus crystal geometry using perturbative SO
interaction inclusion with a Hubbard U correction. They found that
this correction improves a comparison of theoretical and experimental
MAE and shown that MAE is strongly affected by $a$. However, their
absolute values of MAE are much smaller than the experimental
ones. Also, the spin reorientation transition in their calculations
occurs at the lattice constants corresponding to approximately 500
K. The differences are probably related to the different treatment of
the relativistic effects (for Bi-based systems it can be important),
different values of a Hubbard parameters, and different LSDA+$U$
schemes.

\section{excited state properties.}

\subsection{Magneto-optical properties.}

In this section we provide a theoretical explanation of the MO
properties of MnBi.

\begin{figure}[tbp!]
\begin{center}
\includegraphics[width=0.99\columnwidth]{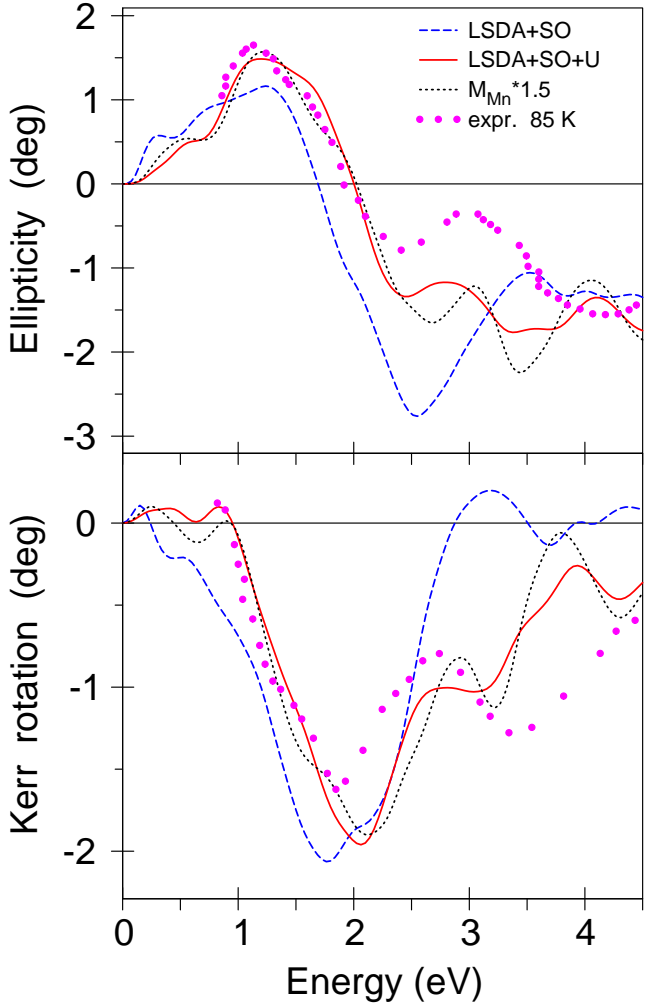}
\end{center}
\caption{\label{MO}(Color online) Calculated in the LSDA+SO (blue
  dashed lines) and LSDA+SO+$U$ (red full lines) approximations polar
  Kerr rotation ($\theta_K$) and Kerr ellipticity ($\varepsilon_K$)
  spectra of MnBi in comparison with the experimental measurements
  from Ref. \onlinecite{DiUc96}. M*1.5 denotes an exchange splitting
  of 150\% of the first-principles value. }
\end{figure}

The experimental Kerr spectra as well as the calculated ones are shown
in Fig. \ref{MO}. The Kerr rotation is denoted by $\theta_K$ and the
Kerr ellipticity by $\varepsilon_K$. First-principles LSDA theory
predicts a very large Kerr rotation in MnBi of about $-2^{\circ}$ at
1.8 eV. This is even larger than the measured peak value of
$-1.6^{\circ}$. \cite{DiUc96} The experiment shows a second maximum in
the Kerr angle at 3.4 eV. Here the LSDA calculations give only a
shoulder. We found that the reason for such disagreement is the
underestimation of the $M_s$ in the LSDA. The $M_s$ at the Mn site is
equal to 3.572 $\mb$ in the LSDA. However, the experimental value at
low temperature is equal to 4.25 $\mb$. \cite{YYC+11} In
Fig. \ref{MO}, we present the calculated Kerr spectra in the
LSDA+SO+$U$ (red full lines) and the LSDA+SO calculations with an
artificially increased exchange splitting on the Mn site by 1.5 times
(black dotted curves). In both latter calculations the $M_s$ is quite
close to the experimental value (4.234 $\mb$ and 4.257 $\mb$ for the
LSDA+SO+$U$ and the LSDA+SO with increased spin splitting,
respectively). Both spectra have similar shapes with much better
reproduction of the second maximum in the Kerr angle at 3.4
eV. Another feature of the experimental Kerr rotation is that it
exhibits a sign reversal at 0.9 eV. This sign reversal is actually
also given by the LSDA+SO method, but for a smaller energy. The
LSDA+SO+$U$ method and the LSDA+SO one with increased spin splitting
perfectly reproduces the energy of a sign reversal at 0.9 eV. The
LSDA+SO+$U$ approximation also reproduces better the observed shape of
the Kerr ellipticity and a sign reversal at around 2 eV (see upper
panel of Fig. \ref{MO}).

\begin{figure}[tbp!]
\begin{center}
\includegraphics[width=0.99\columnwidth]{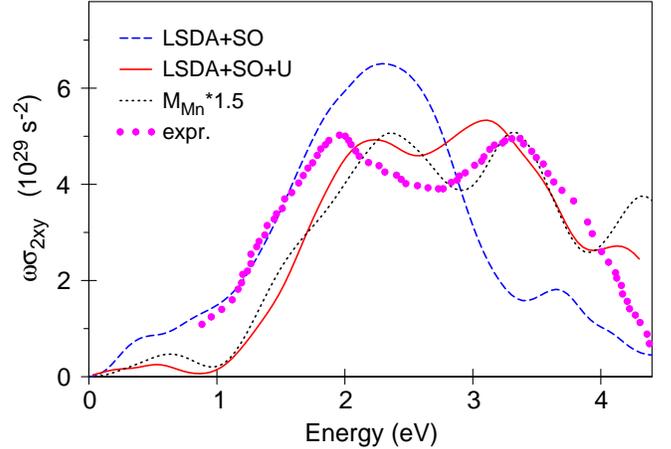}
\end{center}
\caption{\label{Sig_xy}(Color online) Calculated in the LSDA+SO (blue
  dashed lines) and LSDA+SO+$U$ (red full lines) approximations
  off-diagonal component ($\sigma_{2xy}$) of the conductivity tensor
  for MnBi in comparison with the experimental measurements in MnBi
  from Ref. \onlinecite{DiUc96}. M*1.5 denotes an exchange splitting
  of 150\% of the first-principles value }
\end{figure}

The Kerr spectra depend on the MO conductivity spectra in an entangled
way. Therefore, it is difficult to assign features in the Kerr spectra
to particular band transitions. The absorptive parts of the optical
conductivity $\sigma_{1sxx}$ and $\sigma_{2sxy}$ however, relate
directly to the interband optical transitions, and therefore provide
more physical insight. \cite{book:RS90,book:AHY04} The calculated
absorptive part of off-diagonal optical conductivity $\sigma_{2xy}$
for MnBi is shown in Fig. \ref{Sig_xy}. The main peak in the Kerr
rotation of MnBi is due to the maximum in the $\sigma_{2xy}$ at 1.8
eV. The second fine structure in MnBi Kerr spectrum at 3.4 eV
corresponds to the high energy peak in the $\sigma_{2xy}$ at the same
energy. The LSDA+SO calculations strongly underestimate the intensity
of the second high energy peak in the $\sigma_{2xy}$. As a result of
this, the LSDA+SO fails to correctly describe the second negative peak
in the Kerr rotation at 3.4 eV. On the other hand, the LSDA+SO+$U$ and
the LSDA+SO with increased spin splitting quite well reproduce the
intensity of the second high energy peak in the $\sigma_{2xy}$
spectrum and therefore better describes the 3.4 eV peak in the Kerr
rotation.

We can conclude that the main reason for a failure of LDSA to describe
the MO properties in MnBi is the significant underestimation of spin
magnetic moment on the Mn atom. Two very different techniques (the
LSDA+$U$ method and the application of external magnetic field on the
Mn atom) produced similar spin moment enhancement and consequently
better MO values.

\begin{figure}[tbp!]
\begin{center}
\includegraphics[width=0.99\columnwidth]{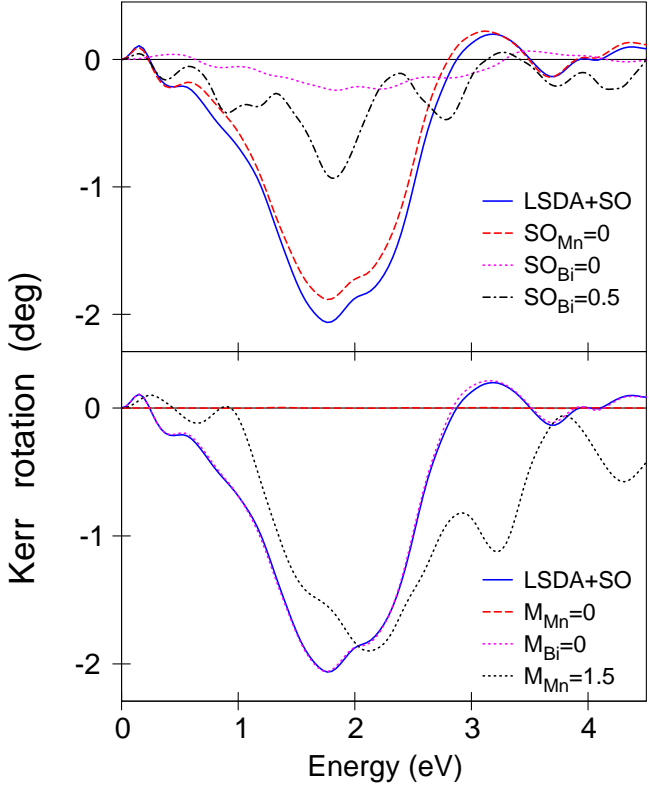}
\end{center}
\caption{\label{MO_M_SO}(Color online) The effect of scaling of the
  exchange splitting (upper panel) and value of SO constant (lower
  panel) on either Mn or Bi atoms on the MO Kerr spectra in MnBi (see
  the text). }
\end{figure}

It is important to identify the origin of the large Kerr effect in
MnBi. To this end, we examine the dependence of the MO spectra on the
exchange splitting and the SO interaction. The exchange splitting and
the SO coupling are studied by scaling the corresponding terms in the
Hamiltonian artificially with a constant prefactor. These
modifications can be done within each atomic sphere independently, so
that we can investigate the separate effects of these quantities on Mn
and on Bi. The outcomes of these model calculations for the Kerr
rotation of MnBi are shown in Fig. \ref{MO_M_SO}. In the lower panel,
the importance of the exchange splitting is illustrated. When the
exchange splitting on Bi is set to zero, the Kerr rotation remains as
it is. But when we do the same for the exchange splitting on Mn, the
Kerr rotation totally vanishes. This implies that the exchange
splitting due to Mn is crucial for the sizable Kerr rotation, but that
of Bi is not important. Furthermore, an enhancement of the exchange
splitting on Mn by a factor of 1.5 (dotted line) leads to a much
larger peak in the Kerr rotation at 3.4 eV. The upper panel of
Fig. \ref{MO_M_SO} shows the dependence on the SO coupling. If we set
the SO coupling on Mn to zero, the Kerr rotation does not change very
much (dashed red line). On the other hand, when the SO coupling on Bi
is zero, the Kerr rotation almost disappears (dotted magenta
line). Thus, the SO coupling of Bi is equally responsible for the
large Kerr rotation as is the exchange splitting of Mn. An
intermediate scaling of the SO coupling of Bi by a factor of 0.5 leads
to an approximately half as large Kerr angle, thereby illustrating the
almost linear dependence of the Kerr effect on the SO interaction of
Bi in this compound.

\subsection{X-ray magnetic circular dichroism.}

Motivated by the developing interest in obtaining element specific
magnetic moment information provided by XMCD measurements. We
calculate the XAS and XMCD spectra of MnBi at the Mn $K$, and $L_{3}$
and at the Bi $M_{2,3}$, $M_{4,5}$, $N_{2,3}$, $N_{4,5}$, $N_{6,7}$ ,
and $O_{2,3}$ edges.

\begin{figure}[tbp!]
\begin{center}
\includegraphics[width=0.99\columnwidth]{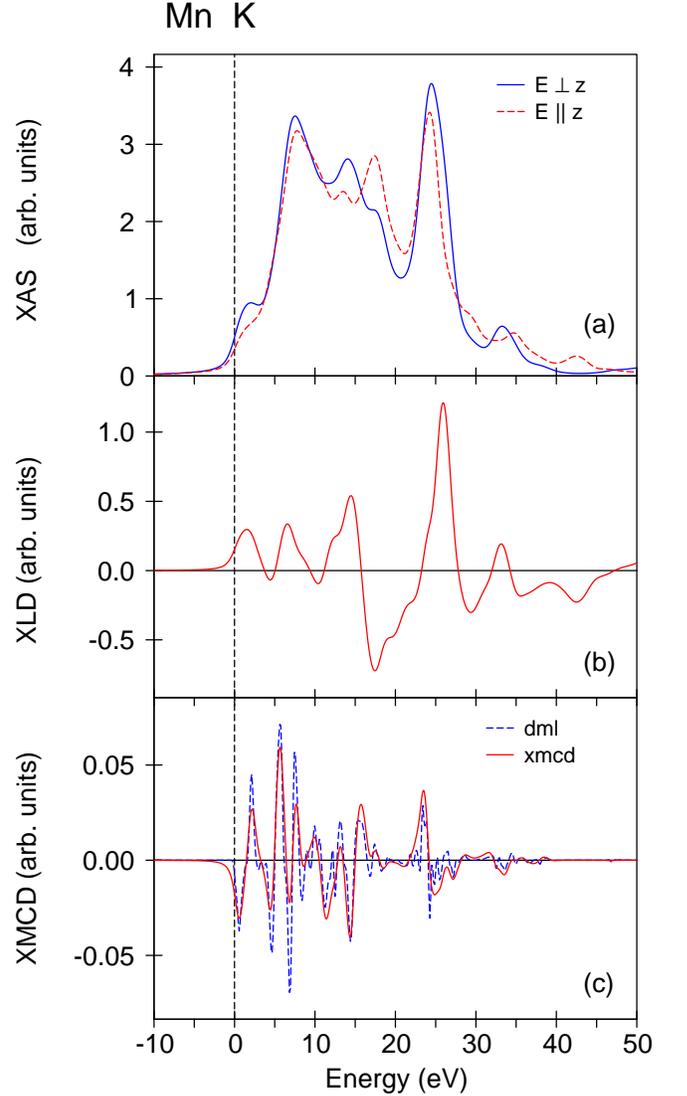}
\end{center}
\caption{\label{Mn_K}(Color online) (a) the theoretically calculated
  x-ray absorption spectra of MnBi at the Mn $K$ edge with the
  electric field vector of the x-rays parallel (red dashed curve) and
  perpendicular (blue full curve) to the z-axis; (b) theoretically
  calculated XLD spectra at the Mn $K$ edge; (c) the theoretically
  calculated XMCD spectrum at the Mn $K$ edge (red full curve) and
  $dm_l$ function (blue dashed curve). The calculations have been done
  using the LSDA+SO+$U$ approach.}
\end{figure}

Figure \ref{Mn_K}(a) shows the theoretically calculated x-ray
absorption spectra at the Mn $K$ edge in MnBi with the electric field
vector of the x-rays both parallel (dashed red curve) and
perpendicular (full blue curve) to the $c$ axis. The associated XLD
signal (obtained by taking the difference of the XA spectra for the
two polarizations) is given in the panel (b) of
Fig. \ref{Mn_K}. Fig. \ref{Mn_K}(c) shows the theoretically calculated
XMCD in terms of the difference in absorption $\Delta \mu_{\rm K} =
\mu^+_{\rm K} - \mu^-_{\rm K}$ for left and right circularly polarized
radiation in MnBi. After comparing the panels (b) and (c), we can
conclude that the XMCD signal is almost one order of magnitude smaller
than the corresponding XLD signal. Also, the spectra have major peaks
in different energy intervals. Major peaks in the XMCD spectrum are
mostly located in the 0 to 15 eV energy interval. However, the XML
spectrum possesses the major peaks above 15 eV.

Because dipole allowed transitions dominate the absorption spectrum
for unpolarized radiation, the absorption coefficient $\mu^0_{\rm
  K}(E)$ reflects primarily the DOS of unoccupied 4$p$-like states
$N_p(E)$ of Mn above the Fermi level. Due to the energy dependent
radial matrix element for the $1s \to 4p$, there is no strict
one-to-one correspondence between $\mu_{\rm K}(E)$ and $N_p(E)$. The
exchange splitting of the initial $1s$-core state is extremely small
and therefore only the exchange and SO splitting of the final
4$p$-states is responsible for the observed dichroism at the
$K$-edge. For this reason, the dichroism is found to be quite small
(lower panel of Fig. \ref{Mn_K}).

To illustrate the influence of SO interaction on the final states
involved in the transitions, let us introduce a site-dependent
function $dm_{tl}(E)$ given by: \cite{UUA+00a}

\begin{equation}
dm_{tl}(E) =  \sum_{m_j}  \sum_{n \bf k}
\langle \Psi_{tl}^{n\bf k} | \hat{l}_z | \Psi_{jm_j} \rangle
\delta (E - E_{n \bf k})\, ,
\label{dml_e}
\end{equation}
where $\hat{l}_z$ is the $z$--projection of the angular momentum
operator, $E_{n \bf k}$ and $\Psi_{tl}^{n \bf k}$ are the energy of
the $n$-th band and the part of the corresponding LMTO wave function
formed by the states with the angular momentum $l$ inside the atomic
sphere centered at the site $t$, respectively. In analogy to the
$l$-projected density of states, $dm_{tl}(E)$ can be referred to as
site- and $l$-projected density of the expectation value of
$\hat{l}_z$.

The $4p - 3d$ hybridization and the SO interaction in the 4$p$ states
play a crucial role for the Mn $K$ edge dichroism.  As seen in
Fig.\ \ref{Mn_K}(c), the $K$ XMCD spectrum and $dm_{tl}(E)$ functions
are closely related to one another and give a rather simple and
straightforward interpretation of the XMCD spectra at the $K$ edge.

Because of the dipole selection rules and apart from the
4$s_{1/2}$-states (which have a small contribution to the XAS due to
relatively small 2$p$ $\to$ 4$s$ matrix elements \cite{book:AHY04}),
only 3$d_{3/2}$-states occur as final states for $L_2$ XAS for
unpolarized radiation. Whereas for $L_3$ XAS the 3$d_{5/2}$-states
also contribute. Although the 2$p_{3/2}$ $\to$ 3$d_{3/2}$ radial
matrix elements are only slightly smaller than elements for the
2$p_{3/2}$ $\to$ 3$d_{5/2}$ transitions the angular matrix elements
strongly suppress the 2$p_{3/2}$ $\to$ 3$d_{3/2}$
contribution. Therefore in neglecting the energy dependence of the
radial matrix elements, the $L_2$- and the $L_3$-spectrum can be
viewed as a direct mapping of the DOS curve for 3$d_{3/2}$- and
3$d_{5/2}$-character, respectively.

\begin{figure}[tbp!]
\begin{center}
\includegraphics[width=0.99\columnwidth]{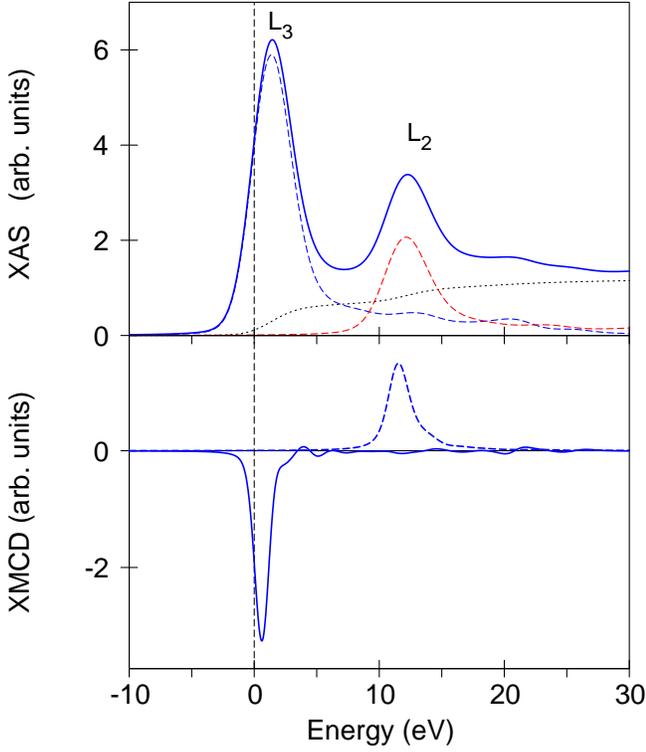}
\end{center}
\caption{\label{Mn_L23}(Color online) X-ray absorption (top panel) and
  XMCD spectra (lower panel) at the Mn $L_{2,3}$ edges calculated using
  the LSDA+SO+$U$. }
\end{figure}

In contrast to the $K$-edge, the dichroism at the $L_{2}$- and
$L_{3}$-edges is also influenced by the SO coupling of the initial
2$p$-core states. This gives rise to a very pronounced dichroism in
comparison with the dichroism at the $K$ edge.  Fig. \ref{Mn_L23}
shows the theoretically calculated Mn $L_{2,3}$ XMCD spectra in
MnBi. The XMCD spectra at the $L_{2,3}$-edges are mostly determined by
the strength of the SO coupling of the initial 2$p$-core states and
spin-polarization of the final empty 3$d_{3/2,5/2}$ states. The
exchange splitting of the 2$p$-core states as well as the SO coupling
of the 3$d$-valence states are of minor importance for the XMCD at the
$L_{2,3}$-edge of 3$d$- transition metals. \cite{book:AHY04}

As mentioned above, XMCD investigations supply information on magnetic
properties in a component resolved way. This seems especially
interesting if there is a magnetic moment induced at a normally
non-magnetic element by neighboring magnetic atoms. The underlying
mechanism of the magnetic and MO properties of the systems considered
here is the well known ability of transition metals to induce large
spin polarization of Bi via strong 3$d$-6$p$ hybridization and
exchange interaction.

\begin{figure}[tbp!]
\begin{center}
\includegraphics[width=0.45\columnwidth]{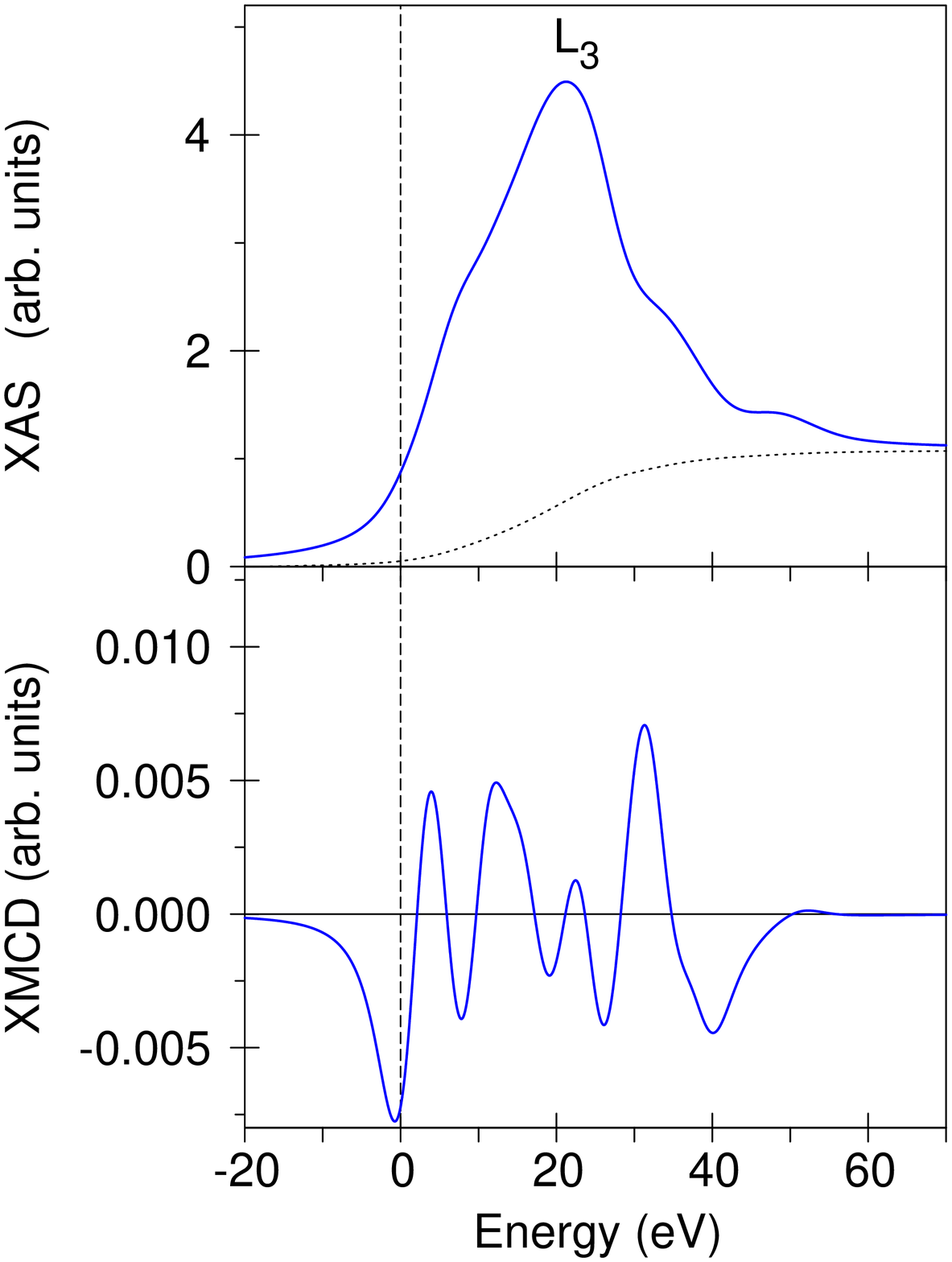}
\includegraphics[width=0.35\columnwidth]{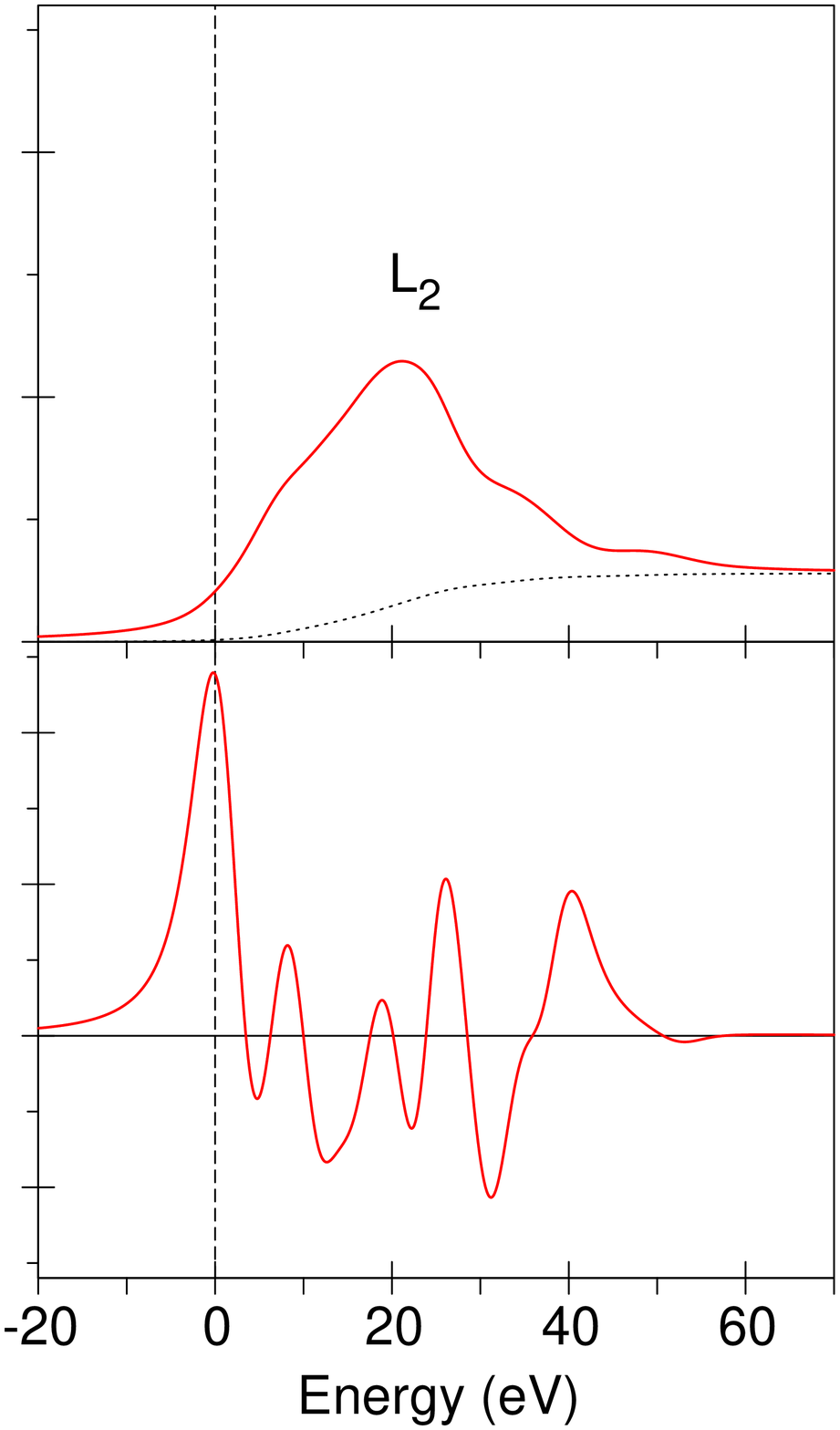}
\end{center}
\caption{\label{Bi_L23}(Color online) X-ray absorption (top panel) and
  XMCD spectra (lower panel) at the Bi $L_{2,3}$ edges calculated using
  the LSDA+SO+$U$. }
\end{figure}

Results of the theoretical calculations for the circular dichroism at
the $L_{2,3}$-edge of Bi are shown in Figure \ref{Bi_L23}. The XMCD
spectrum is negative at the $L_3$ edge and positive at the $L_2$ edge
as has been seen for the XMCD spectra at the $L_{2,3}$-edges of Mn
(Fig. \ref{Mn_L23}). The XMCD in Bi at the $L_3$ and $L_2$ edges are
of nearly equal magnitude.  This suggests that an orbital magnetic
moment almost vanishes in Bi 5$d$ states in MnBi.

\begin{figure}[tbp!]
\begin{center}
\includegraphics[width=0.48\columnwidth]{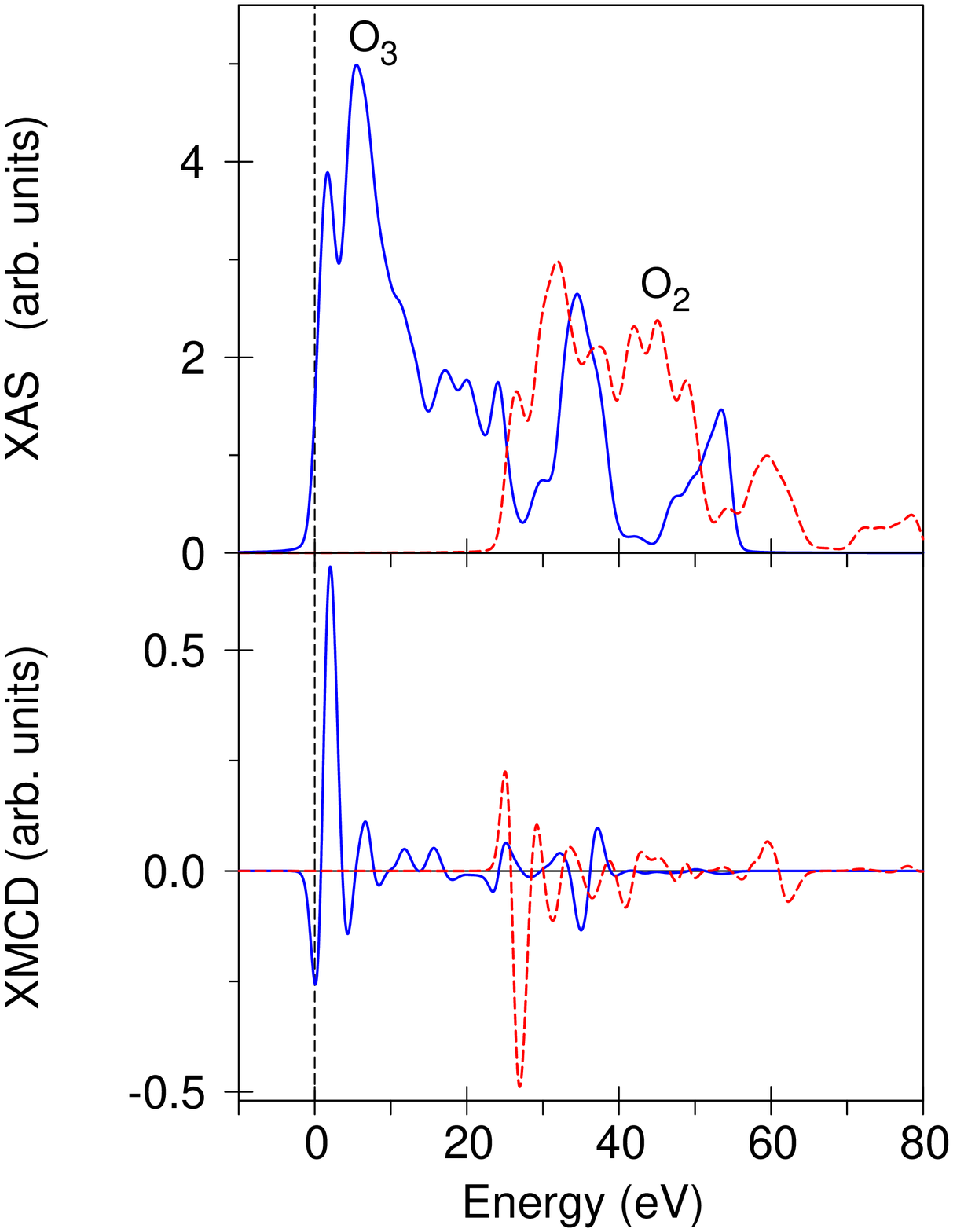}
\includegraphics[width=0.41\columnwidth]{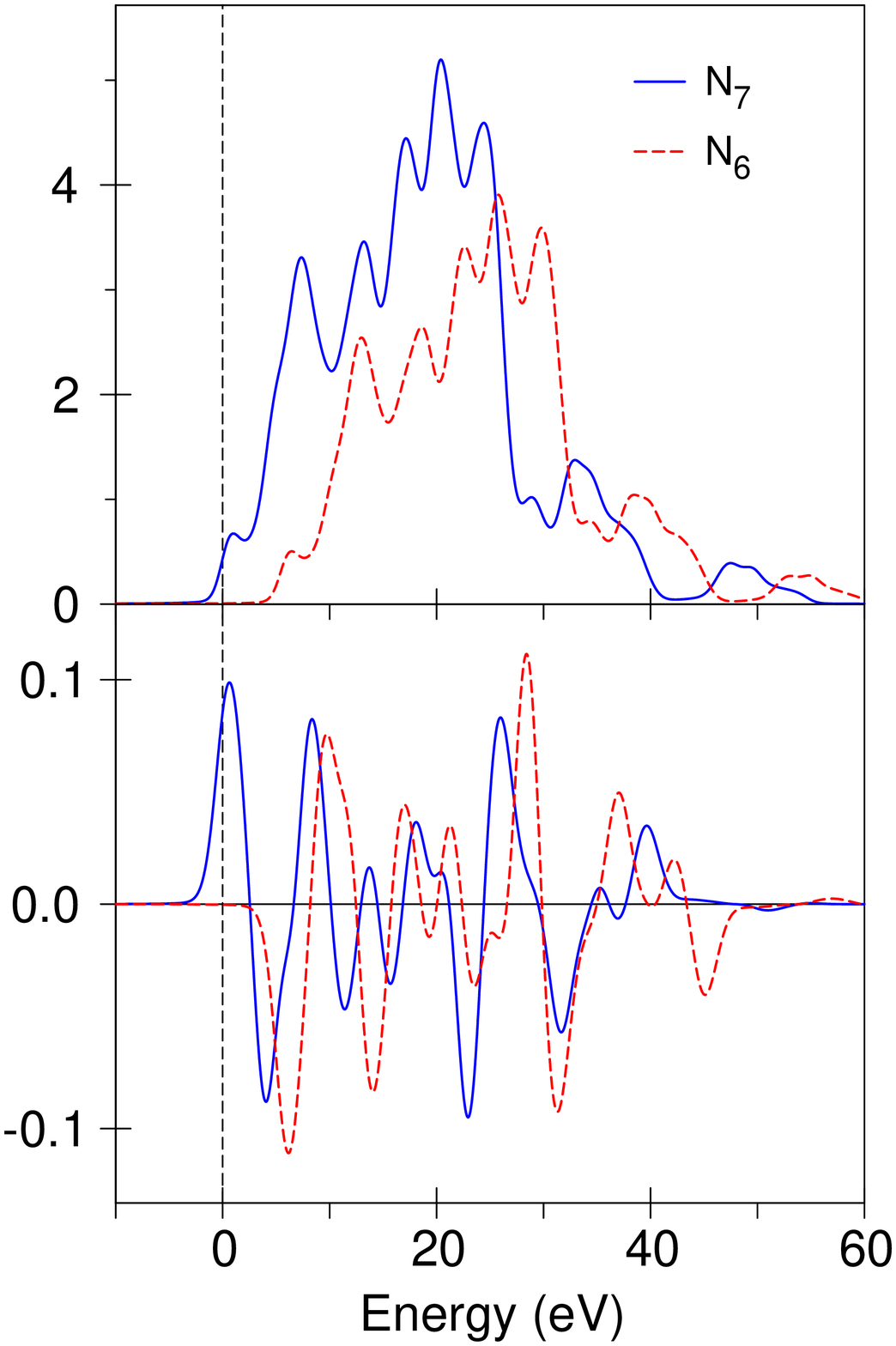}
\end{center}\caption{\label{Bi_O23_N67}(Color online) X-ray absorption
 (top panels) and XMCD (lower panels) spectra at the Bi $O_{2,3}$
  (left panels) and $N_{6,7}$ (right panels) edges calculated in the
  LSDA+SO+$U$. }
\end{figure}

To investigate the influence of the initial state on the resulting Bi
XMCD spectra, we also calculated the XAS and XMCD spectra of MnBi at
the $M_{2,3}$, $M_{4,5}$, $N_{2,3}$, $N_{4,5}$, $N_{6,7}$ , and
$O_{2,3}$ edges. We found a systematic decreasing of the XMCD spectra
in terms of $R=\Delta \mu/(2 \mu^0_K)$ in the row
$L_{2,3}-M_{2,3}-N_{2,3}$ edges. Although the magnetic dichroism of
quasi-core states ($O_{2,3}$ and $N_{6,7}$ edges) became almost one
order of magnitude larger as it was at the $L_{2,3}$ edges (compare
Figs. \ref{Bi_L23} and \ref{Bi_O23_N67}). Besides, the lifetime widths
of the core $O_{2,3}$ and $N_{6,7}$ levels are much smaller than the
$L_{2,3}$ values. \cite{book:FI92} The spectroscopy of Bi atoms in the
ultra-soft x-ray energy range at the $O_{2,3}$ and $N_{6,7}$ edges may
therefore be a useful tool for investigating the electronic and
magnetic structure of MnBi.

\begin{figure}[tbp!]
\begin{center}
\includegraphics[width=0.46\columnwidth]{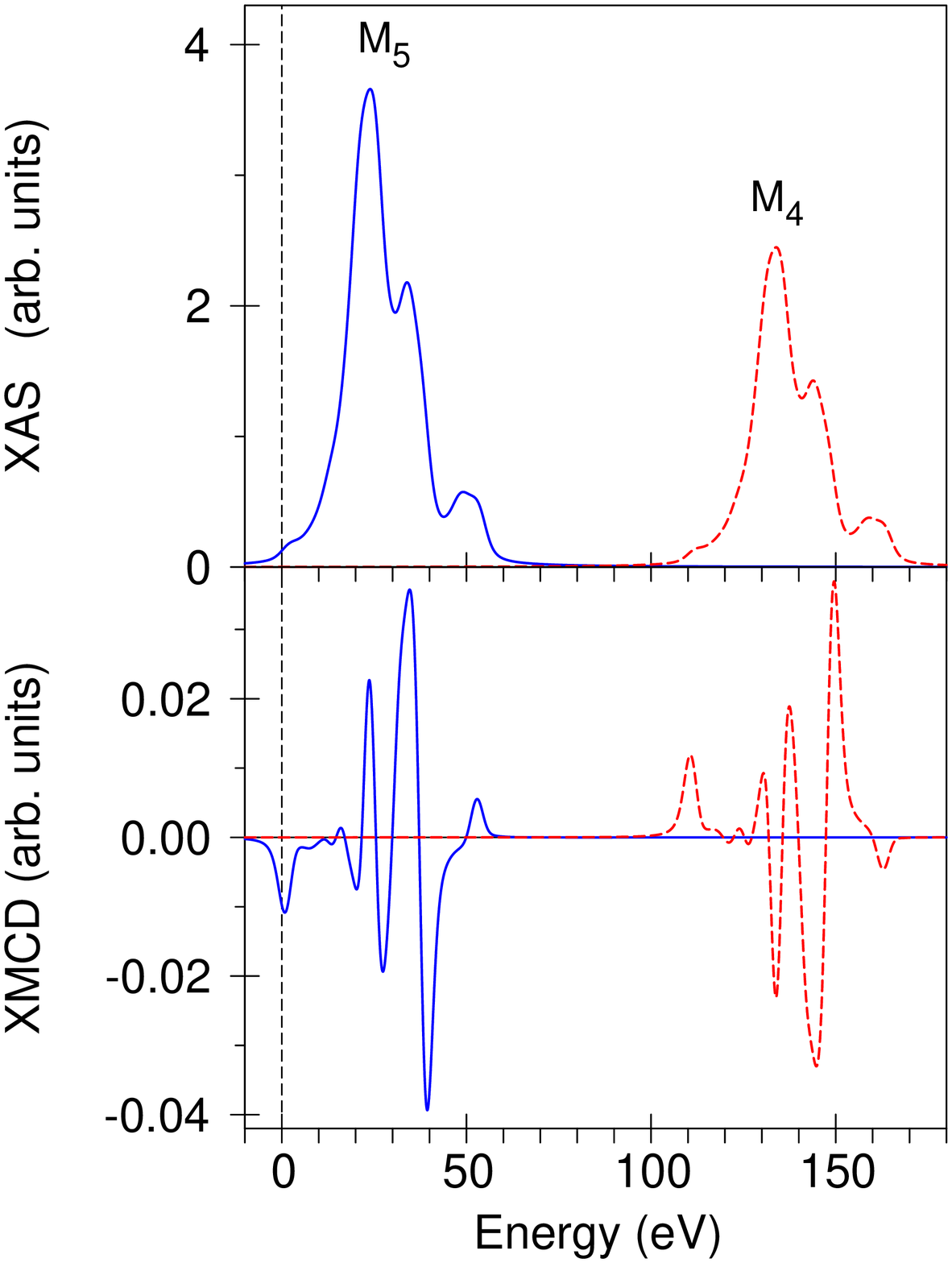}
\includegraphics[width=0.41\columnwidth]{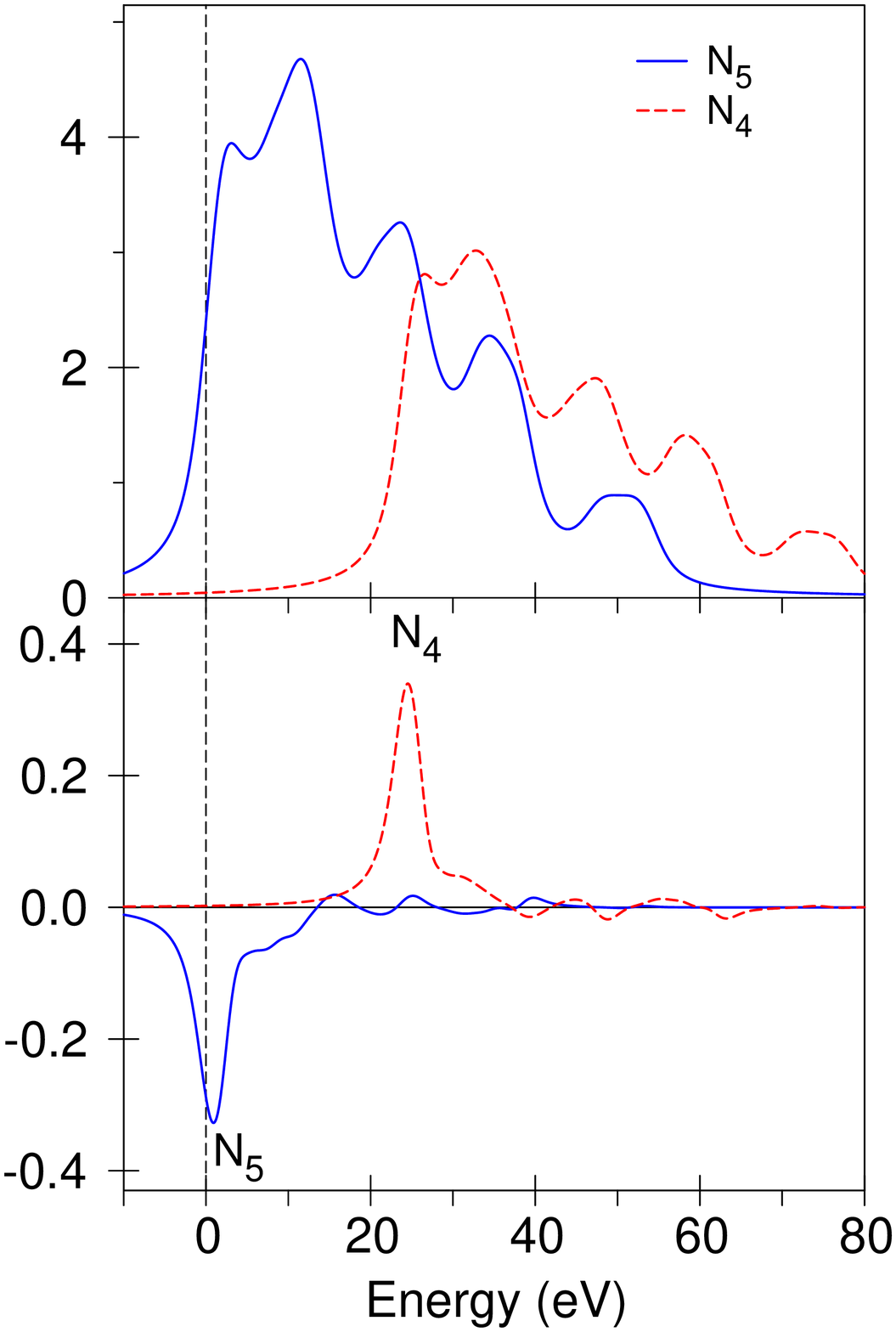}
\end{center}
\caption{\label{Bi_M45_N45}(Color online) X-ray absorption (top
  panels) and XMCD spectra (lower panels) at the Bi $M_{4,5}$ (left
  panels) and edges $N_{4,5}$ (right panels) calculated using the
  LSDA+SO+$U$. }
\end{figure}

Bi $M_{4,5}$ and $N_{4,5}$ spectra may be considered as an analog of
the $K$ spectrum to some extent. The $K$ absorption spectrum reflects
the energy distribution of empty $p_{1/2}$ and $p_{3/2}$ energy
states. The $M_4$ ($N_4$) absorption spectrum is due to the dipole
selection rules occur during the transition from the 3$d_{3/2}$
(4$d_{3/2}$) core states to the $p_{1/2}$, $p_{3/2}$, and $f_{5/2}$
valence states above the Fermi level. The $p_{3/2}$, $f_{5/2}$, and
$f_{7/2}$ states contribute to the $M_5$ ($N_5$) XASs. Results of the
theoretical calculations of the circular dichroism in absorption at
the $M_{4,5}$ and $N_{4,5}$ edges of MnBi are shown in Figure
\ref{Bi_M45_N45}.

It is interesting to compare the Bi XAS and XMCD spectra at the
$L_{2,3}$, $O_{2,3}$ and $N_{6,7}$ edges. Due to the dipole selection
rules, for unpolarized radiation (apart from the $s_{1/2}$ states
which have a small contribution to the XAS) only the 3$d_{3/2}$ states
occur as final states for the $L_2$ as well as for the $O_2$
spectra. The $L_3$ and $O_3$ spectra reflect the energy distribution
of both the 3$d_{3/2}$ and 3$d_{5/2}$ empty states. On the other hand,
the $N_7$ absorption spectrum reflects only the 3$d_{5/2}$ states (the
density of the $g_{7/2,9/2}$ states is really very small), whereas for
the $N_6$, XAS both the 3$d_{3/2}$ and 3$d_{5/2}$ states
contribute. We therefore have an inverse situation: the $N_6$
absorption spectra correspond to the $L_3$ and $O_3$ spectra, and the
$N_7$ is the analog of the $L_2$ and $O_2$ ones. This situation is
clearly seen in Figs. \ref{Bi_O23_N67} and \ref{Bi_L23} where the
theoretically calculated XMCD spectra of MnBi at the $O_{2,3}$,
$N_{6,7}$, and $L_{2,3}$ edges is presented. The XMCD spectra at $L_3$
edges are almost identical to the spectra at the $N_6$ edges. The XMCD
spectra at the $L_2$ edges are also very similar to the spectra at the
$N_7$ edges (but not identical because the energy distribution of the
Bi 5$d_{3/2}$ and 5$d_{5/2}$ states is not exactly the same due to SO
interaction). The spectral shape of the magnetic circular dichroism at
the $O_{3}$ ($O_{2}$) edge also resembles the corresponding dichroism
at the $L_{3}$ and $N_{6}$ ($L_{2}$ and $N_{7}$) edges. However, the
intensity of the XMCD signal at the $O_{3}$ ($O_{2}$) edge is
relatively larger near the edge.  One can argue that at least for Bi
the $L_{2,3}$ and $N_{6,7}$, the spectra predominantly reflect atomic
aspects of the valence bands. For the $O_{2,3}$ edges, the itinerant
aspects are more important.

Because of the relatively small SO splitting of the 4$f$ states of Bi
($\sim$3.0 eV), the $N_6$ and the $N_7$ spectra have an appreciable
overlap. For this reason, the $N_7$ spectrum contributes to some
extent to the structure of the total $N_{6,7}$ spectrum in the region
of the $N_6$ edge (see Fig. \ref{Bi_O23_N67}). To decompose a
corresponding experimental $N_{6,7}$ spectrum into its $N_6$ and $N_7$
parts will therefore be quite difficult in general. A similar problem
also occurs in the XAS spectra at the $O$ edge (see
Fig. \ref{Bi_O23_N67}).

\section{\bigskip Conclusions.}

We have performed relativistic LSDA+SO+$U$ calculations of the
electronic structure and Fermi surface properties in MnBi. It was
found the SO interaction and Coulomb correlations strongly affects the
shape of the FS by changing the size and even the topology of some
sheets of the FS in MnBi.

The x-ray absorption and x-ray magnetic dichroism of MnBi at the Mn
$K$, $L_{2,3}$, and $M_{2,3}$ edges and the Bi $L_{2,3}$, $M_{2,3}$,
$M_{4,5}$, $N_{2,3}$, $N_{4,5}$, $N_{6,7}$, and $O_{2,3}$ edges have
been studied.

We showed that the SO coupling of Bi, the exchange splitting of Mn,
and the Mn-Bi hybridization are all crucial components for a large MO
Kerr effect in MnBi. It was determined that the main reason for a
failure of previous LSDA calculations to describe the MO properties
in MnBi is the significant underestimation of spin magnetic moment on
the Mn atom. Two very different techniques (the LSDA+$U$ method and an
application of external magnetic field on the Mn atom) produce similar
spin moment enhancement and consequently better MO and MAE values.

The LSDA+SO+$U$ approach provides a proper value of magnetic moment on
Mn atoms and correct easy magnetization direction for the entire
temperature interval. The LSDA+SO+$U$ theory produces MAE which is in
very good quantitative agreement with experimental results in the 150
K to 450 K temperature range. The physical reason for the observed
spin reorientation is the unusual dependence of anisotropic Bi-Bi
exchange on in-plane lattice parameter or exchange striction
mechanism.

It appears that the spin reorientation in MnBi correlates with the
inversion of orbital moment minimum on Bi atom with temperature
according to the third Hund's rule. At lower temperature a maximum of
orbital moment corresponds to a minimum of the total energy, while at
room temperature it corresponds to the energy maximum.

For a large amount of properties that have been discussed in this
paper, no prior theoretical or experimental studies have been
performed. With high quality single crystal samples already available,
the opportunity to perform the initial experimental study of many
properties discussed above becomes feasible and therefore out other
theoretical predictions can be tested.

\section*{Acknowledgments}

This research is supported in part by the U.S. Department
of Energy (DOE) ARPA-E (REACT 0472-1526); the Critical
Materials Institute, an Energy Innovation Hub funded by the
U.S. DOE and by the Office of Basic Energy Science, Division of
Materials Science and Engineering. Ames Laboratory is operated for the
U.S. DOE by Iowa State University under contract DE-AC02-07CH11358.
V.P.A. is thankful to S.V.Antropov for useful comments.
V.N.A. gratefully acknowledges the hospitality at Ames Laboratory
during his stay there.


\begin{thebibliography}{94}
\expandafter\ifx\csname natexlab\endcsname\relax\def\natexlab#1{#1}\fi
\expandafter\ifx\csname bibnamefont\endcsname\relax
  \def\bibnamefont#1{#1}\fi
\expandafter\ifx\csname bibfnamefont\endcsname\relax
  \def\bibfnamefont#1{#1}\fi
\expandafter\ifx\csname citenamefont\endcsname\relax
  \def\citenamefont#1{#1}\fi
\expandafter\ifx\csname url\endcsname\relax
  \def\url#1{\texttt{#1}}\fi
\expandafter\ifx\csname urlprefix\endcsname\relax\def\urlprefix{URL }\fi
\providecommand{\bibinfo}[2]{#2}
\providecommand{\eprint}[2][]{\url{#2}}

\bibitem[{\citenamefont{Yang et~al.}(2001)\citenamefont{Yang, Kamaraju, Yelon,
  and James}}]{YKY+01}
\bibinfo{author}{\bibfnamefont{J.}~\bibnamefont{Yang}},
  \bibinfo{author}{\bibfnamefont{K.}~\bibnamefont{Kamaraju}},
  \bibinfo{author}{\bibfnamefont{W.}~\bibnamefont{Yelon}}, \bibnamefont{and}
  \bibinfo{author}{\bibfnamefont{W.}~\bibnamefont{James}},
  \bibinfo{journal}{Appl. Physics Lett.} \textbf{\bibinfo{volume}{79}},
  \bibinfo{pages}{1846} (\bibinfo{year}{2001}).

\bibitem[{\citenamefont{Di et~al.}(1992)\citenamefont{Di, Iwata, Tsunashima,
  and Uchiyama}}]{DIT+92}
\bibinfo{author}{\bibfnamefont{G.~Q.} \bibnamefont{Di}},
  \bibinfo{author}{\bibfnamefont{S.}~\bibnamefont{Iwata}},
  \bibinfo{author}{\bibfnamefont{S.}~\bibnamefont{Tsunashima}},
  \bibnamefont{and} \bibinfo{author}{\bibfnamefont{S.}~\bibnamefont{Uchiyama}},
  \bibinfo{journal}{J. Magn. Magn. Mater.} \textbf{\bibinfo{volume}{104}},
  \bibinfo{pages}{1023} (\bibinfo{year}{1992}).

\bibitem[{\citenamefont{Heikes}(1955)}]{Heikes55}
\bibinfo{author}{\bibfnamefont{R.~R.} \bibnamefont{Heikes}},
  \bibinfo{journal}{Phys. Rev.} \textbf{\bibinfo{volume}{99}},
  \bibinfo{pages}{446} (\bibinfo{year}{1955}).

\bibitem[{\citenamefont{R\"udiger and G\"untherodt}(2000)}]{RuGu00}
\bibinfo{author}{\bibfnamefont{U.}~\bibnamefont{R\"udiger}} \bibnamefont{and}
  \bibinfo{author}{\bibfnamefont{G.}~\bibnamefont{G\"untherodt}},
  \bibinfo{journal}{J. Appl. Phys.} \textbf{\bibinfo{volume}{88}},
  \bibinfo{pages}{4221} (\bibinfo{year}{2000}).

\bibitem[{\citenamefont{Yang et~al.}(2002)\citenamefont{Yang, Yelon, James,
  Cai, Kornecki, Roy, Ali, and I'Heritier}}]{YYJC+02}
\bibinfo{author}{\bibfnamefont{J.~B.} \bibnamefont{Yang}},
  \bibinfo{author}{\bibfnamefont{W.~B.} \bibnamefont{Yelon}},
  \bibinfo{author}{\bibfnamefont{W.~J.} \bibnamefont{James}},
  \bibinfo{author}{\bibfnamefont{Q.}~\bibnamefont{Cai}},
  \bibinfo{author}{\bibfnamefont{M.}~\bibnamefont{Kornecki}},
  \bibinfo{author}{\bibfnamefont{S.}~\bibnamefont{Roy}},
  \bibinfo{author}{\bibfnamefont{N.}~\bibnamefont{Ali}}, \bibnamefont{and}
  \bibinfo{author}{\bibfnamefont{P.}~\bibnamefont{I'Heritier}},
  \bibinfo{journal}{J. Phys.: Condens. Matter} \textbf{\bibinfo{volume}{14}},
  \bibinfo{pages}{6509} (\bibinfo{year}{2002}).

\bibitem[{\citenamefont{Chen and Stutius}(1974)}]{ChSt74}
\bibinfo{author}{\bibfnamefont{T.}~\bibnamefont{Chen}} \bibnamefont{and}
  \bibinfo{author}{\bibfnamefont{W.}~\bibnamefont{Stutius}},
  \bibinfo{journal}{IEEE Trans. Magn.} \textbf{\bibinfo{volume}{10}},
  \bibinfo{pages}{581} (\bibinfo{year}{1974}).

\bibitem[{\citenamefont{Andersen et~al.}(1967)\citenamefont{Andersen, H\"alg,
  Fischer, and Stoll}}]{AHF+67}
\bibinfo{author}{\bibfnamefont{A.~F.} \bibnamefont{Andersen}},
  \bibinfo{author}{\bibfnamefont{W.}~\bibnamefont{H\"alg}},
  \bibinfo{author}{\bibfnamefont{P.}~\bibnamefont{Fischer}}, \bibnamefont{and}
  \bibinfo{author}{\bibfnamefont{E.}~\bibnamefont{Stoll}},
  \bibinfo{journal}{Acta. Chem. Scand.} \textbf{\bibinfo{volume}{21}},
  \bibinfo{pages}{1543} (\bibinfo{year}{1967}).

\bibitem[{\citenamefont{Heusler}(1904)}]{Heusler04}
\bibinfo{author}{\bibfnamefont{F.}~\bibnamefont{Heusler}}, \bibinfo{journal}{Z.
  Angew. Chern.} \textbf{\bibinfo{volume}{17}}, \bibinfo{pages}{260}
  (\bibinfo{year}{1904}).

\bibitem[{\citenamefont{Bekier}(1914)}]{Bek14}
\bibinfo{author}{\bibfnamefont{E.}~\bibnamefont{Bekier}},
  \bibinfo{journal}{Intern. Z. Metallographie} \textbf{\bibinfo{volume}{7}},
  \bibinfo{pages}{83} (\bibinfo{year}{1914}).

\bibitem[{\citenamefont{Parravano and Perret}(1915)}]{PaPe15}
\bibinfo{author}{\bibfnamefont{N.}~\bibnamefont{Parravano}} \bibnamefont{and}
  \bibinfo{author}{\bibfnamefont{U.}~\bibnamefont{Perret}},
  \bibinfo{journal}{Gazz. Chern. Ita.} \textbf{\bibinfo{volume}{451}},
  \bibinfo{pages}{390} (\bibinfo{year}{1915}).

\bibitem[{\citenamefont{Hilpert and Dieckmann}(1911)}]{HiHa11}
\bibinfo{author}{\bibfnamefont{S.}~\bibnamefont{Hilpert}} \bibnamefont{and}
  \bibinfo{author}{\bibfnamefont{T.}~\bibnamefont{Dieckmann}},
  \bibinfo{journal}{Ber. Deut. Chern. Ges.} \textbf{\bibinfo{volume}{44}},
  \bibinfo{pages}{2831} (\bibinfo{year}{1911}).

\bibitem[{\citenamefont{Furst and Halla}(1938)}]{FuHa38}
\bibinfo{author}{\bibfnamefont{U.}~\bibnamefont{Furst}} \bibnamefont{and}
  \bibinfo{author}{\bibfnamefont{F.}~\bibnamefont{Halla}}, \bibinfo{journal}{Z.
  Physik Chern. B} \textbf{\bibinfo{volume}{40}}, \bibinfo{pages}{285}
  (\bibinfo{year}{1938}).

\bibitem[{\citenamefont{Montignie}(1938)}]{Mon38}
\bibinfo{author}{\bibfnamefont{E.}~\bibnamefont{Montignie}},
  \bibinfo{journal}{Bull. Soc. Chim.} \textbf{\bibinfo{volume}{5}},
  \bibinfo{pages}{343} (\bibinfo{year}{1938}).

\bibitem[{\citenamefont{Halla and Montignie}(1939)}]{HaMO39}
\bibinfo{author}{\bibfnamefont{F.}~\bibnamefont{Halla}} \bibnamefont{and}
  \bibinfo{author}{\bibfnamefont{E.}~\bibnamefont{Montignie}},
  \bibinfo{journal}{Z. Physik Chern. B} \textbf{\bibinfo{volume}{42}},
  \bibinfo{pages}{153} (\bibinfo{year}{1939}).

\bibitem[{\citenamefont{Guillaud}(1943)}]{book:Guill43}
\bibinfo{author}{\bibfnamefont{C.}~\bibnamefont{Guillaud}},
  \emph{\bibinfo{title}{PhD Thesis}} (\bibinfo{publisher}{University of
  Strasbourg}, \bibinfo{address}{Strasbourg, France}, \bibinfo{year}{1943}).

\bibitem[{\citenamefont{Adams}(1953)}]{Adams53}
\bibinfo{author}{\bibfnamefont{E.}~\bibnamefont{Adams}}, \bibinfo{journal}{Rev.
  Mod. Phys.} \textbf{\bibinfo{volume}{25}}, \bibinfo{pages}{306}
  (\bibinfo{year}{1953}).

\bibitem[{\citenamefont{Hihara and K\"oi}(1970)}]{HiKo70}
\bibinfo{author}{\bibfnamefont{T.}~\bibnamefont{Hihara}} \bibnamefont{and}
  \bibinfo{author}{\bibfnamefont{Y.}~\bibnamefont{K\"oi}}, \bibinfo{journal}{J.
  Phys. Soc. Jpn.} \textbf{\bibinfo{volume}{29}}, \bibinfo{pages}{343}
  (\bibinfo{year}{1970}).

\bibitem[{\citenamefont{Stutius et~al.}(1974)\citenamefont{Stutius, Chen, and
  Sandin}}]{SCS74}
\bibinfo{author}{\bibfnamefont{W.~E.} \bibnamefont{Stutius}},
  \bibinfo{author}{\bibfnamefont{T.}~\bibnamefont{Chen}}, \bibnamefont{and}
  \bibinfo{author}{\bibfnamefont{T.~R.} \bibnamefont{Sandin}},
  \bibinfo{journal}{AIP Conf. Proc.} \textbf{\bibinfo{volume}{18}},
  \bibinfo{pages}{1222} (\bibinfo{year}{1974}).

\bibitem[{\citenamefont{Williams et~al.}(1957)\citenamefont{Williams, Sherwood,
  Foster, and Keller}}]{WSF+57}
\bibinfo{author}{\bibfnamefont{H.~J.} \bibnamefont{Williams}},
  \bibinfo{author}{\bibfnamefont{R.~C.} \bibnamefont{Sherwood}},
  \bibinfo{author}{\bibfnamefont{F.~G.} \bibnamefont{Foster}},
  \bibnamefont{and} \bibinfo{author}{\bibfnamefont{E.~M.}
  \bibnamefont{Keller}}, \bibinfo{journal}{J. Appl. Phys.}
  \textbf{\bibinfo{volume}{28}}, \bibinfo{pages}{1181} (\bibinfo{year}{1957}).

\bibitem[{\citenamefont{Mayer}(1958)}]{May58}
\bibinfo{author}{\bibfnamefont{L.}~\bibnamefont{Mayer}}, \bibinfo{journal}{J.
  Appl. Phys.} \textbf{\bibinfo{volume}{29}}, \bibinfo{pages}{1003}
  (\bibinfo{year}{1958}).

\bibitem[{\citenamefont{Mayer}(1960)}]{May60}
\bibinfo{author}{\bibfnamefont{L.}~\bibnamefont{Mayer}}, \bibinfo{journal}{J.
  Appl. Phys.} \textbf{\bibinfo{volume}{31}}, \bibinfo{pages}{3845}
  (\bibinfo{year}{1960}).

\bibitem[{\citenamefont{Chen}(1971)}]{Chen71}
\bibinfo{author}{\bibfnamefont{D.}~\bibnamefont{Chen}}, \bibinfo{journal}{J.
  Appl. Phys.} \textbf{\bibinfo{volume}{42}}, \bibinfo{pages}{3625}
  (\bibinfo{year}{1971}).

\bibitem[{\citenamefont{Buschow}(1988)}]{book:Bus88}
\bibinfo{author}{\bibfnamefont{K.~H.~J.} \bibnamefont{Buschow}}, in
  \emph{\bibinfo{booktitle}{Ferromagnetic Materials}}, edited by
  \bibinfo{editor}{\bibfnamefont{E.~P.} \bibnamefont{Wohlfarth}}
  \bibnamefont{and} \bibinfo{editor}{\bibfnamefont{K.~H.~J.}
  \bibnamefont{Buschow}} (\bibinfo{publisher}{North-Holland},
  \bibinfo{address}{Amsterdam}, \bibinfo{year}{1988}),
  vol.~\bibinfo{volume}{4}, p. \bibinfo{pages}{588}.

\bibitem[{\citenamefont{Huang et~al.}(1993)\citenamefont{Huang, Zheng, Luo,
  Yang, and Wang}}]{HZL+93}
\bibinfo{author}{\bibfnamefont{D.}~\bibnamefont{Huang}},
  \bibinfo{author}{\bibfnamefont{X.~W.} \bibnamefont{Zheng}},
  \bibinfo{author}{\bibfnamefont{C.~P.} \bibnamefont{Luo}},
  \bibinfo{author}{\bibfnamefont{H.~S.} \bibnamefont{Yang}}, \bibnamefont{and}
  \bibinfo{author}{\bibfnamefont{Y.~J.} \bibnamefont{Wang}},
  \bibinfo{journal}{J. Appl. Phys.} \textbf{\bibinfo{volume}{75}},
  \bibinfo{pages}{6351} (\bibinfo{year}{1993}).

\bibitem[{\citenamefont{Di and Uchiyama}(1996)}]{DiUc96}
\bibinfo{author}{\bibfnamefont{G.~Q.} \bibnamefont{Di}} \bibnamefont{and}
  \bibinfo{author}{\bibfnamefont{S.}~\bibnamefont{Uchiyama}},
  \bibinfo{journal}{Phys. Rev. B} \textbf{\bibinfo{volume}{53}},
  \bibinfo{pages}{3327} (\bibinfo{year}{1996}).

\bibitem[{\citenamefont{Harder et~al.}(1998)\citenamefont{Harder, Menzel,
  Widmer, and Schoenes}}]{HMW+98}
\bibinfo{author}{\bibfnamefont{K.}~\bibnamefont{Harder}},
  \bibinfo{author}{\bibfnamefont{D.}~\bibnamefont{Menzel}},
  \bibinfo{author}{\bibfnamefont{T.}~\bibnamefont{Widmer}}, \bibnamefont{and}
  \bibinfo{author}{\bibfnamefont{J.}~\bibnamefont{Schoenes}},
  \bibinfo{journal}{J. Appl. Phys.} \textbf{\bibinfo{volume}{84}},
  \bibinfo{pages}{3625} (\bibinfo{year}{1998}).

\bibitem[{\citenamefont{Coehoorn and de~Groot}(1985)}]{CoGr85}
\bibinfo{author}{\bibfnamefont{R.}~\bibnamefont{Coehoorn}} \bibnamefont{and}
  \bibinfo{author}{\bibfnamefont{R.}~\bibnamefont{de~Groot}},
  \bibinfo{journal}{J. Phys. F: Met. Phys.} \textbf{\bibinfo{volume}{15}},
  \bibinfo{pages}{2135} (\bibinfo{year}{1985}).

\bibitem[{\citenamefont{Wang and Huang}(1993)}]{WaHu93}
\bibinfo{author}{\bibfnamefont{Z.~W.} \bibnamefont{Wang}} \bibnamefont{and}
  \bibinfo{author}{\bibfnamefont{M.~C.} \bibnamefont{Huang}},
  \bibinfo{journal}{Chinese Phys. Lett.} \textbf{\bibinfo{volume}{10}},
  \bibinfo{pages}{612} (\bibinfo{year}{1993}).

\bibitem[{\citenamefont{Sabiryanov and Jaswal}(1996)}]{SaJa96}
\bibinfo{author}{\bibfnamefont{R.~F.} \bibnamefont{Sabiryanov}}
  \bibnamefont{and} \bibinfo{author}{\bibfnamefont{S.~S.}
  \bibnamefont{Jaswal}}, \bibinfo{journal}{Phys. Rev. B}
  \textbf{\bibinfo{volume}{53}}, \bibinfo{pages}{313} (\bibinfo{year}{1996}).

\bibitem[{\citenamefont{Oppeneer et~al.}(1996)\citenamefont{Oppeneer, Antonov,
  Kraft, Eschrig, Yaresko, and Perlov}}]{OAK+96a}
\bibinfo{author}{\bibfnamefont{P.~M.} \bibnamefont{Oppeneer}},
  \bibinfo{author}{\bibfnamefont{V.~N.} \bibnamefont{Antonov}},
  \bibinfo{author}{\bibfnamefont{T.}~\bibnamefont{Kraft}},
  \bibinfo{author}{\bibfnamefont{H.}~\bibnamefont{Eschrig}},
  \bibinfo{author}{\bibfnamefont{A.~N.} \bibnamefont{Yaresko}},
  \bibnamefont{and} \bibinfo{author}{\bibfnamefont{A.~Y.}
  \bibnamefont{Perlov}}, \bibinfo{journal}{J. Appl. Phys.}
  \textbf{\bibinfo{volume}{80}}, \bibinfo{pages}{1099} (\bibinfo{year}{1996}).

\bibitem[{\citenamefont{Uspenskii et~al.}(1996)\citenamefont{Uspenskii,
  Kulatov, and Halilov}}]{UKH96}
\bibinfo{author}{\bibfnamefont{A.~Y.~A.} \bibnamefont{Uspenskii}},
  \bibinfo{author}{\bibfnamefont{E.~T.} \bibnamefont{Kulatov}},
  \bibnamefont{and} \bibinfo{author}{\bibfnamefont{S.~V.}
  \bibnamefont{Halilov}}, \textbf{\bibinfo{volume}{241}}, \bibinfo{pages}{89}
  (\bibinfo{year}{1996}).

\bibitem[{\citenamefont{K\"ohler and K\"ubler}(1996)}]{KoKu96}
\bibinfo{author}{\bibfnamefont{J.}~\bibnamefont{K\"ohler}} \bibnamefont{and}
  \bibinfo{author}{\bibfnamefont{J.}~\bibnamefont{K\"ubler}},
  \bibinfo{journal}{J. Phys.: Condens. Matter} \textbf{\bibinfo{volume}{8}},
  \bibinfo{pages}{8681} (\bibinfo{year}{1996}).

\bibitem[{\citenamefont{K\"ohler and K\"ubler}(1997)}]{KoKu97}
\bibinfo{author}{\bibfnamefont{J.}~\bibnamefont{K\"ohler}} \bibnamefont{and}
  \bibinfo{author}{\bibfnamefont{J.}~\bibnamefont{K\"ubler}},
  \bibinfo{journal}{Physica B} \textbf{\bibinfo{volume}{402}},
  \bibinfo{pages}{237} (\bibinfo{year}{1997}).

\bibitem[{\citenamefont{Burkova et~al.}(1999)\citenamefont{Burkova,
  Ovchinnikov, Seredkin, and Yakovchuk}}]{BOS+99}
\bibinfo{author}{\bibfnamefont{L.~V.} \bibnamefont{Burkova}},
  \bibinfo{author}{\bibfnamefont{S.~G.} \bibnamefont{Ovchinnikov}},
  \bibinfo{author}{\bibfnamefont{V.~A.} \bibnamefont{Seredkin}},
  \bibnamefont{and} \bibinfo{author}{\bibfnamefont{V.~Y.}
  \bibnamefont{Yakovchuk}}, \bibinfo{journal}{J. Magn. Magn. Mater.}
  \textbf{\bibinfo{volume}{195}}, \bibinfo{pages}{531} (\bibinfo{year}{1999}).

\bibitem[{\citenamefont{Ravindran et~al.}(1999)\citenamefont{Ravindran, Delin,
  James, Johansson, Wills, Ahuja, and Eriksson}}]{RDJ+99}
\bibinfo{author}{\bibfnamefont{P.}~\bibnamefont{Ravindran}},
  \bibinfo{author}{\bibfnamefont{A.}~\bibnamefont{Delin}},
  \bibinfo{author}{\bibfnamefont{P.}~\bibnamefont{James}},
  \bibinfo{author}{\bibfnamefont{B.}~\bibnamefont{Johansson}},
  \bibinfo{author}{\bibfnamefont{J.~M.} \bibnamefont{Wills}},
  \bibinfo{author}{\bibfnamefont{R.}~\bibnamefont{Ahuja}}, \bibnamefont{and}
  \bibinfo{author}{\bibfnamefont{O.}~\bibnamefont{Eriksson}},
  \bibinfo{journal}{Phys. Rev. B} \textbf{\bibinfo{volume}{59}},
  \bibinfo{pages}{15680} (\bibinfo{year}{1999}).

\bibitem[{\citenamefont{Tan et~al.}(2000)\citenamefont{Tan, Tao, and
  Bao}}]{TTB00}
\bibinfo{author}{\bibfnamefont{M.~Q.} \bibnamefont{Tan}},
  \bibinfo{author}{\bibfnamefont{X.~M.} \bibnamefont{Tao}}, \bibnamefont{and}
  \bibinfo{author}{\bibfnamefont{S.~N.} \bibnamefont{Bao}},
  \bibinfo{journal}{Chinese Phys.} \textbf{\bibinfo{volume}{9}},
  \bibinfo{pages}{55} (\bibinfo{year}{2000}).

\bibitem[{\citenamefont{Mavropoulos et~al.}(2004)\citenamefont{Mavropoulos,
  Sato, Zeller, Dederichs, Popescu, and Ebert}}]{MSZ+04}
\bibinfo{author}{\bibfnamefont{P.}~\bibnamefont{Mavropoulos}},
  \bibinfo{author}{\bibfnamefont{K.}~\bibnamefont{Sato}},
  \bibinfo{author}{\bibfnamefont{R.}~\bibnamefont{Zeller}},
  \bibinfo{author}{\bibfnamefont{P.~H.} \bibnamefont{Dederichs}},
  \bibinfo{author}{\bibfnamefont{V.}~\bibnamefont{Popescu}}, \bibnamefont{and}
  \bibinfo{author}{\bibfnamefont{H.}~\bibnamefont{Ebert}},
  \bibinfo{journal}{Phys. Rev. B} \textbf{\bibinfo{volume}{69}},
  \bibinfo{pages}{054424} (\bibinfo{year}{2004}).

\bibitem[{\citenamefont{Zheng and Davenport}(2004)}]{ZhDa04}
\bibinfo{author}{\bibfnamefont{J.~C.} \bibnamefont{Zheng}} \bibnamefont{and}
  \bibinfo{author}{\bibfnamefont{J.~W.} \bibnamefont{Davenport}},
  \bibinfo{journal}{Phys. Rev. B} \textbf{\bibinfo{volume}{69}},
  \bibinfo{pages}{144415} (\bibinfo{year}{2004}).

\bibitem[{\citenamefont{Weng et~al.}(2006)\citenamefont{Weng, Kawazoe, and
  Dong}}]{WKD+06}
\bibinfo{author}{\bibfnamefont{H.}~\bibnamefont{Weng}},
  \bibinfo{author}{\bibfnamefont{Y.}~\bibnamefont{Kawazoe}}, \bibnamefont{and}
  \bibinfo{author}{\bibfnamefont{J.}~\bibnamefont{Dong}},
  \bibinfo{journal}{Phys. Rev. B} \textbf{\bibinfo{volume}{74}},
  \bibinfo{pages}{085205} (\bibinfo{year}{2006}).

\bibitem[{\citenamefont{Li et~al.}(2007)\citenamefont{Li, Ariizumi, Koyanagi,
  and Suzuki}}]{LAK+07}
\bibinfo{author}{\bibfnamefont{M.}~\bibnamefont{Li}},
  \bibinfo{author}{\bibfnamefont{T.}~\bibnamefont{Ariizumi}},
  \bibinfo{author}{\bibfnamefont{K.}~\bibnamefont{Koyanagi}}, \bibnamefont{and}
  \bibinfo{author}{\bibfnamefont{S.}~\bibnamefont{Suzuki}},
  \bibinfo{journal}{Jap. J. Appl. Phys.} \textbf{\bibinfo{volume}{46}},
  \bibinfo{pages}{3455} (\bibinfo{year}{2007}).

\bibitem[{\citenamefont{Sanyal and Eriksson}(2008)}]{SaEr08}
\bibinfo{author}{\bibfnamefont{B.}~\bibnamefont{Sanyal}} \bibnamefont{and}
  \bibinfo{author}{\bibfnamefont{O.}~\bibnamefont{Eriksson}},
  \bibinfo{journal}{J. Appl. Phys.} \textbf{\bibinfo{volume}{103}},
  \bibinfo{pages}{07D704} (\bibinfo{year}{2008}).

\bibitem[{\citenamefont{Kahal and Ferhat}(2010)}]{KaFe10}
\bibinfo{author}{\bibfnamefont{L.}~\bibnamefont{Kahal}} \bibnamefont{and}
  \bibinfo{author}{\bibfnamefont{M.}~\bibnamefont{Ferhat}},
  \bibinfo{journal}{J. Appl. Phys.} \textbf{\bibinfo{volume}{107}},
  \bibinfo{pages}{043910} (\bibinfo{year}{2010}).

\bibitem[{\citenamefont{Goncalves et~al.}(2011)\citenamefont{Goncalves, Amaral,
  Correia, and Lopes}}]{GAC+11}
\bibinfo{author}{\bibfnamefont{J.~N.} \bibnamefont{Goncalves}},
  \bibinfo{author}{\bibfnamefont{V.~S.} \bibnamefont{Amaral}},
  \bibinfo{author}{\bibfnamefont{J.~G.} \bibnamefont{Correia}},
  \bibnamefont{and} \bibinfo{author}{\bibfnamefont{A.~M.~L.}
  \bibnamefont{Lopes}}, \bibinfo{journal}{Phys. Rev. B}
  \textbf{\bibinfo{volume}{83}}, \bibinfo{pages}{104421}
  (\bibinfo{year}{2011}).

\bibitem[{\citenamefont{Kharel et~al.}(2011)\citenamefont{Kharel, Thapa,
  Lukashev, Sabirianov, Tsymbal, Sellmyer, and Nadgorny}}]{KTL+11}
\bibinfo{author}{\bibfnamefont{P.}~\bibnamefont{Kharel}},
  \bibinfo{author}{\bibfnamefont{P.}~\bibnamefont{Thapa}},
  \bibinfo{author}{\bibfnamefont{P.}~\bibnamefont{Lukashev}},
  \bibinfo{author}{\bibfnamefont{R.~F.} \bibnamefont{Sabirianov}},
  \bibinfo{author}{\bibfnamefont{E.~Y.} \bibnamefont{Tsymbal}},
  \bibinfo{author}{\bibfnamefont{D.~J.} \bibnamefont{Sellmyer}},
  \bibnamefont{and} \bibinfo{author}{\bibfnamefont{B.}~\bibnamefont{Nadgorny}},
  \bibinfo{journal}{Phys. Rev. B} \textbf{\bibinfo{volume}{83}},
  \bibinfo{pages}{024415} (\bibinfo{year}{2011}).

\bibitem[{\citenamefont{Bandaru et~al.}(1998)\citenamefont{Bandaru, Sands,
  Kubota, and Marinero}}]{BSK+98}
\bibinfo{author}{\bibfnamefont{P.~R.} \bibnamefont{Bandaru}},
  \bibinfo{author}{\bibfnamefont{T.~D.} \bibnamefont{Sands}},
  \bibinfo{author}{\bibfnamefont{Y.}~\bibnamefont{Kubota}}, \bibnamefont{and}
  \bibinfo{author}{\bibfnamefont{E.~E.} \bibnamefont{Marinero}},
  \bibinfo{journal}{Appl. Physics Lett.} \textbf{\bibinfo{volume}{72}},
  \bibinfo{pages}{2337} (\bibinfo{year}{1998}).

\bibitem[{\citenamefont{Schoenes}(1992)}]{book:Sch92}
\bibinfo{author}{\bibfnamefont{J.}~\bibnamefont{Schoenes}}, in
  \emph{\bibinfo{booktitle}{Electronic and Magnetic Properties of Metals and
  Ceramics}}, edited by \bibinfo{editor}{\bibfnamefont{R.~W.}
  \bibnamefont{Cahn}},
  \bibinfo{editor}{\bibfnamefont{P.}~\bibnamefont{Haasen}}, \bibnamefont{and}
  \bibinfo{editor}{\bibfnamefont{E.~J.} \bibnamefont{Kramer}}
  (\bibinfo{publisher}{Verlag Chemie}, \bibinfo{address}{Weinheim},
  \bibinfo{year}{1992}), vol.~\bibinfo{volume}{3A} of
  \emph{\bibinfo{series}{Materials Science and Technology}}, p.
  \bibinfo{pages}{147}, \bibinfo{note}{volume editor: K. H. J. Buschow}.

\bibitem[{\citenamefont{Kubo}(1957)}]{Kub57}
\bibinfo{author}{\bibfnamefont{R.}~\bibnamefont{Kubo}}, \bibinfo{journal}{J.
  Phys. Soc. Jpn.} \textbf{\bibinfo{volume}{12}}, \bibinfo{pages}{570}
  (\bibinfo{year}{1957}).

\bibitem[{\citenamefont{Wang and Callaway}(1974)}]{WC74}
\bibinfo{author}{\bibfnamefont{C.~S.} \bibnamefont{Wang}} \bibnamefont{and}
  \bibinfo{author}{\bibfnamefont{J.}~\bibnamefont{Callaway}},
  \bibinfo{journal}{Phys. Rev. B} \textbf{\bibinfo{volume}{9}},
  \bibinfo{pages}{4897} (\bibinfo{year}{1974}).

\bibitem[{\citenamefont{Antonov et~al.}(1993)\citenamefont{Antonov, Bagljuk,
  Perlov, Nemoshkalenko, Antonov, Andersen, and Jepsen}}]{ABP+93}
\bibinfo{author}{\bibfnamefont{V.~N.} \bibnamefont{Antonov}},
  \bibinfo{author}{\bibfnamefont{A.~I.} \bibnamefont{Bagljuk}},
  \bibinfo{author}{\bibfnamefont{A.~Y.} \bibnamefont{Perlov}},
  \bibinfo{author}{\bibfnamefont{V.~V.} \bibnamefont{Nemoshkalenko}},
  \bibinfo{author}{\bibfnamefont{V.~N.} \bibnamefont{Antonov}},
  \bibinfo{author}{\bibfnamefont{O.~K.} \bibnamefont{Andersen}},
  \bibnamefont{and} \bibinfo{author}{\bibfnamefont{O.}~\bibnamefont{Jepsen}},
  \bibinfo{journal}{Low Temp. Phys.} \textbf{\bibinfo{volume}{19}},
  \bibinfo{pages}{494} (\bibinfo{year}{1993}).

\bibitem[{\citenamefont{Antonov et~al.}(2004)\citenamefont{Antonov, Harmon, and
  Yaresko}}]{book:AHY04}
\bibinfo{author}{\bibfnamefont{V.}~\bibnamefont{Antonov}},
  \bibinfo{author}{\bibfnamefont{B.}~\bibnamefont{Harmon}}, \bibnamefont{and}
  \bibinfo{author}{\bibfnamefont{A.}~\bibnamefont{Yaresko}},
  \emph{\bibinfo{title}{Electronic Structure and Magneto-Optical Properties of
  Solids}} (\bibinfo{publisher}{Kluwer}, \bibinfo{address}{Dordrecht},
  \bibinfo{year}{2004}).

\bibitem[{\citenamefont{Kunes and Oppeneer}(2003)}]{KuOp03}
\bibinfo{author}{\bibfnamefont{J.}~\bibnamefont{Kunes}} \bibnamefont{and}
  \bibinfo{author}{\bibfnamefont{P.~M.} \bibnamefont{Oppeneer}},
  \bibinfo{journal}{Phys. Rev. B} \textbf{\bibinfo{volume}{67}},
  \bibinfo{pages}{024431} (\bibinfo{year}{2003}).

\bibitem[{\citenamefont{Smith and Wjin}(1959)}]{book:SW59}
\bibinfo{author}{\bibfnamefont{J.}~\bibnamefont{Smith}} \bibnamefont{and}
  \bibinfo{author}{\bibfnamefont{H.~P.~J.} \bibnamefont{Wjin}},
  \emph{\bibinfo{title}{Ferrites}} (\bibinfo{publisher}{Philips Technical
  Library}, \bibinfo{address}{Eindhoven}, \bibinfo{year}{1959}).

\bibitem[{\citenamefont{Antonov et~al.}(1995)\citenamefont{Antonov, Perlov,
  Shpak, and Yaresko}}]{APS+95}
\bibinfo{author}{\bibfnamefont{V.~N.} \bibnamefont{Antonov}},
  \bibinfo{author}{\bibfnamefont{A.~Y.} \bibnamefont{Perlov}},
  \bibinfo{author}{\bibfnamefont{A.~P.} \bibnamefont{Shpak}}, \bibnamefont{and}
  \bibinfo{author}{\bibfnamefont{A.~N.} \bibnamefont{Yaresko}},
  \bibinfo{journal}{J. Magn. Magn. Mater.} \textbf{\bibinfo{volume}{146}},
  \bibinfo{pages}{205} (\bibinfo{year}{1995}).

\bibitem[{\citenamefont{Andersen}(1975)}]{And75}
\bibinfo{author}{\bibfnamefont{O.~K.} \bibnamefont{Andersen}},
  \bibinfo{journal}{Phys. Rev. B} \textbf{\bibinfo{volume}{12}},
  \bibinfo{pages}{3060} (\bibinfo{year}{1975}).

\bibitem[{\citenamefont{Bl\"ochl et~al.}(1994)\citenamefont{Bl\"ochl, Jepsen,
  and Andersen}}]{BJA94}
\bibinfo{author}{\bibfnamefont{P.~E.} \bibnamefont{Bl\"ochl}},
  \bibinfo{author}{\bibfnamefont{O.}~\bibnamefont{Jepsen}}, \bibnamefont{and}
  \bibinfo{author}{\bibfnamefont{O.~K.} \bibnamefont{Andersen}},
  \bibinfo{journal}{Phys. Rev. B} \textbf{\bibinfo{volume}{49}},
  \bibinfo{pages}{16223} (\bibinfo{year}{1994}).

\bibitem[{\citenamefont{Perdew and Zunger}(1981)}]{PZ81}
\bibinfo{author}{\bibfnamefont{J.~P.} \bibnamefont{Perdew}} \bibnamefont{and}
  \bibinfo{author}{\bibfnamefont{A.}~\bibnamefont{Zunger}},
  \bibinfo{journal}{Phys. Rev. B} \textbf{\bibinfo{volume}{23}},
  \bibinfo{pages}{5048} (\bibinfo{year}{1981}).

\bibitem[{\citenamefont{Anisimov et~al.}(1991)\citenamefont{Anisimov, Zaanen,
  and Andersen}}]{AZA91}
\bibinfo{author}{\bibfnamefont{V.~I.} \bibnamefont{Anisimov}},
  \bibinfo{author}{\bibfnamefont{J.}~\bibnamefont{Zaanen}}, \bibnamefont{and}
  \bibinfo{author}{\bibfnamefont{O.~K.} \bibnamefont{Andersen}},
  \bibinfo{journal}{Phys. Rev. B} \textbf{\bibinfo{volume}{44}},
  \bibinfo{pages}{943} (\bibinfo{year}{1991}).

\bibitem[{\citenamefont{Hedin}(1965)}]{Hed65}
\bibinfo{author}{\bibfnamefont{L.}~\bibnamefont{Hedin}},
  \bibinfo{journal}{Phys. Rev.} \textbf{\bibinfo{volume}{139}},
  \bibinfo{pages}{A796} (\bibinfo{year}{1965}).

\bibitem[{\citenamefont{Metzner and Vollhardt}(1989)}]{MV89}
\bibinfo{author}{\bibfnamefont{W.}~\bibnamefont{Metzner}} \bibnamefont{and}
  \bibinfo{author}{\bibfnamefont{D.}~\bibnamefont{Vollhardt}},
  \bibinfo{journal}{Phys. Rev. Lett.} \textbf{\bibinfo{volume}{62}},
  \bibinfo{pages}{324} (\bibinfo{year}{1989}).

\bibitem[{\citenamefont{Pruschke et~al.}(1995)\citenamefont{Pruschke, Jarell,
  and Freericks}}]{PJF95}
\bibinfo{author}{\bibfnamefont{T.}~\bibnamefont{Pruschke}},
  \bibinfo{author}{\bibfnamefont{M.}~\bibnamefont{Jarell}}, \bibnamefont{and}
  \bibinfo{author}{\bibfnamefont{J.~K.} \bibnamefont{Freericks}},
  \bibinfo{journal}{Adv. Phys.} \textbf{\bibinfo{volume}{44}},
  \bibinfo{pages}{187} (\bibinfo{year}{1995}).

\bibitem[{\citenamefont{Georges et~al.}(1996)\citenamefont{Georges, Kotliar,
  Krauth, and Rozenberg}}]{GKK+96}
\bibinfo{author}{\bibfnamefont{A.}~\bibnamefont{Georges}},
  \bibinfo{author}{\bibfnamefont{G.}~\bibnamefont{Kotliar}},
  \bibinfo{author}{\bibfnamefont{W.}~\bibnamefont{Krauth}}, \bibnamefont{and}
  \bibinfo{author}{\bibfnamefont{M.~J.} \bibnamefont{Rozenberg}},
  \bibinfo{journal}{Rev. Mod. Phys.} \textbf{\bibinfo{volume}{68}},
  \bibinfo{pages}{13} (\bibinfo{year}{1996}).

\bibitem[{\citenamefont{Yaresko et~al.}(2003)\citenamefont{Yaresko, Antonov,
  and Fulde}}]{YAF03}
\bibinfo{author}{\bibfnamefont{A.~N.} \bibnamefont{Yaresko}},
  \bibinfo{author}{\bibfnamefont{V.~N.} \bibnamefont{Antonov}},
  \bibnamefont{and} \bibinfo{author}{\bibfnamefont{P.}~\bibnamefont{Fulde}},
  \bibinfo{journal}{Phys. Rev. B} \textbf{\bibinfo{volume}{67}},
  \bibinfo{pages}{155103} (\bibinfo{year}{2003}).

\bibitem[{\citenamefont{Bengone et~al.}(2000)\citenamefont{Bengone, Alouani,
  Bl\"ochl, and Hugel}}]{BABH00}
\bibinfo{author}{\bibfnamefont{O.}~\bibnamefont{Bengone}},
  \bibinfo{author}{\bibfnamefont{M.}~\bibnamefont{Alouani}},
  \bibinfo{author}{\bibfnamefont{P.}~\bibnamefont{Bl\"ochl}}, \bibnamefont{and}
  \bibinfo{author}{\bibfnamefont{J.}~\bibnamefont{Hugel}},
  \bibinfo{journal}{Phys. Rev. B} \textbf{\bibinfo{volume}{62}},
  \bibinfo{pages}{16392} (\bibinfo{year}{2000}).

\bibitem[{\citenamefont{Anisimov and Gunarsson}(1991)}]{AG91}
\bibinfo{author}{\bibfnamefont{V.~I.} \bibnamefont{Anisimov}} \bibnamefont{and}
  \bibinfo{author}{\bibfnamefont{G.}~\bibnamefont{Gunarsson}},
  \bibinfo{journal}{Phys. Rev. B} \textbf{\bibinfo{volume}{43}},
  \bibinfo{pages}{7570} (\bibinfo{year}{1991}).

\bibitem[{\citenamefont{Solovyev et~al.}(1994)\citenamefont{Solovyev,
  Dederichs, and Anisimov}}]{SDA94}
\bibinfo{author}{\bibfnamefont{I.~V.} \bibnamefont{Solovyev}},
  \bibinfo{author}{\bibfnamefont{P.~H.} \bibnamefont{Dederichs}},
  \bibnamefont{and} \bibinfo{author}{\bibfnamefont{V.~I.}
  \bibnamefont{Anisimov}}, \bibinfo{journal}{Phys. Rev. B}
  \textbf{\bibinfo{volume}{50}}, \bibinfo{pages}{16861} (\bibinfo{year}{1994}).

\bibitem[{\citenamefont{Dederichs et~al.}(1984)\citenamefont{Dederichs,
  Bl\"ugel, Zeller, and Akai}}]{DBZ+84}
\bibinfo{author}{\bibfnamefont{P.~H.} \bibnamefont{Dederichs}},
  \bibinfo{author}{\bibfnamefont{S.}~\bibnamefont{Bl\"ugel}},
  \bibinfo{author}{\bibfnamefont{R.}~\bibnamefont{Zeller}}, \bibnamefont{and}
  \bibinfo{author}{\bibfnamefont{H.}~\bibnamefont{Akai}},
  \bibinfo{journal}{Phys. Rev. Lett.} \textbf{\bibinfo{volume}{53}},
  \bibinfo{pages}{2512} (\bibinfo{year}{1984}).

\bibitem[{\citenamefont{Pickett et~al.}(1998)\citenamefont{Pickett, Erwin, and
  Ethridge}}]{PEE98}
\bibinfo{author}{\bibfnamefont{W.~E.} \bibnamefont{Pickett}},
  \bibinfo{author}{\bibfnamefont{S.~C.} \bibnamefont{Erwin}}, \bibnamefont{and}
  \bibinfo{author}{\bibfnamefont{E.~C.} \bibnamefont{Ethridge}},
  \bibinfo{journal}{Phys. Rev. B} \textbf{\bibinfo{volume}{58}},
  \bibinfo{pages}{1201} (\bibinfo{year}{1998}).

\bibitem[{\citenamefont{Cococcioni and de~Gironcoli}(2005)}]{CoGi05}
\bibinfo{author}{\bibfnamefont{M.}~\bibnamefont{Cococcioni}} \bibnamefont{and}
  \bibinfo{author}{\bibfnamefont{S.}~\bibnamefont{de~Gironcoli}},
  \bibinfo{journal}{Phys. Rev. B} \textbf{\bibinfo{volume}{71}},
  \bibinfo{pages}{035105} (\bibinfo{year}{2005}).

\bibitem[{\citenamefont{Nakamura et~al.}(2006)\citenamefont{Nakamura, Arita,
  Yoshimoto, and Tsuneyuki}}]{NAY+06}
\bibinfo{author}{\bibfnamefont{K.}~\bibnamefont{Nakamura}},
  \bibinfo{author}{\bibfnamefont{R.}~\bibnamefont{Arita}},
  \bibinfo{author}{\bibfnamefont{Y.}~\bibnamefont{Yoshimoto}},
  \bibnamefont{and}
  \bibinfo{author}{\bibfnamefont{S.}~\bibnamefont{Tsuneyuki}},
  \bibinfo{journal}{Phys. Rev. B} \textbf{\bibinfo{volume}{74}},
  \bibinfo{pages}{235113} (\bibinfo{year}{2006}).

\bibitem[{\citenamefont{Aryasetiawan et~al.}(2006)\citenamefont{Aryasetiawan,
  Karlsson, Jepsen, and Schonberger}}]{AKJ+06}
\bibinfo{author}{\bibfnamefont{F.}~\bibnamefont{Aryasetiawan}},
  \bibinfo{author}{\bibfnamefont{K.}~\bibnamefont{Karlsson}},
  \bibinfo{author}{\bibfnamefont{O.}~\bibnamefont{Jepsen}}, \bibnamefont{and}
  \bibinfo{author}{\bibfnamefont{U.}~\bibnamefont{Schonberger}},
  \bibinfo{journal}{Phys. Rev. B} \textbf{\bibinfo{volume}{74}},
  \bibinfo{pages}{125106} (\bibinfo{year}{2006}).

\bibitem[{\citenamefont{Antonides et~al.}(1997)\citenamefont{Antonides, Janse,
  and Sawatzky}}]{AJS97}
\bibinfo{author}{\bibfnamefont{E.}~\bibnamefont{Antonides}},
  \bibinfo{author}{\bibfnamefont{E.~C.} \bibnamefont{Janse}}, \bibnamefont{and}
  \bibinfo{author}{\bibfnamefont{G.~A.} \bibnamefont{Sawatzky}},
  \bibinfo{journal}{Phys. Rev. B} \textbf{\bibinfo{volume}{15}},
  \bibinfo{pages}{1669} (\bibinfo{year}{1997}).

\bibitem[{\citenamefont{van~der Marel et~al.}(1984)\citenamefont{van~der Marel,
  Sawatzky, and Hillebrecht}}]{MSH84}
\bibinfo{author}{\bibfnamefont{D.}~\bibnamefont{van~der Marel}},
  \bibinfo{author}{\bibfnamefont{G.~A.} \bibnamefont{Sawatzky}},
  \bibnamefont{and} \bibinfo{author}{\bibfnamefont{F.~U.}
  \bibnamefont{Hillebrecht}}, \bibinfo{journal}{Phys. Rev. B}
  \textbf{\bibinfo{volume}{53}}, \bibinfo{pages}{206} (\bibinfo{year}{1984}).

\bibitem[{\citenamefont{Springer and Aryasetiawan}(1998)}]{SpAr98}
\bibinfo{author}{\bibfnamefont{M.}~\bibnamefont{Springer}} \bibnamefont{and}
  \bibinfo{author}{\bibfnamefont{F.}~\bibnamefont{Aryasetiawan}},
  \bibinfo{journal}{Phys. Rev. B} \textbf{\bibinfo{volume}{57}},
  \bibinfo{pages}{4364} (\bibinfo{year}{1998}).

\bibitem[{\citenamefont{Kotani}(2000)}]{Kot00}
\bibinfo{author}{\bibfnamefont{T.}~\bibnamefont{Kotani}}, \bibinfo{journal}{J.
  Phys.: Condens. Matter} \textbf{\bibinfo{volume}{12}}, \bibinfo{pages}{2413}
  (\bibinfo{year}{2000}).

\bibitem[{\citenamefont{Aryasetiawan et~al.}(2004)\citenamefont{Aryasetiawan,
  Imada, Georges, Kotliar, Biermann, and Lichtenstein}}]{AIG+04}
\bibinfo{author}{\bibfnamefont{F.}~\bibnamefont{Aryasetiawan}},
  \bibinfo{author}{\bibfnamefont{M.}~\bibnamefont{Imada}},
  \bibinfo{author}{\bibfnamefont{A.}~\bibnamefont{Georges}},
  \bibinfo{author}{\bibfnamefont{G.}~\bibnamefont{Kotliar}},
  \bibinfo{author}{\bibfnamefont{S.}~\bibnamefont{Biermann}}, \bibnamefont{and}
  \bibinfo{author}{\bibfnamefont{A.~I.} \bibnamefont{Lichtenstein}},
  \bibinfo{journal}{Phys. Rev. B} \textbf{\bibinfo{volume}{70}},
  \bibinfo{pages}{195104} (\bibinfo{year}{2004}).

\bibitem[{\citenamefont{Solovyev and Imada}(2005)}]{SoIm05}
\bibinfo{author}{\bibfnamefont{I.~V.} \bibnamefont{Solovyev}} \bibnamefont{and}
  \bibinfo{author}{\bibfnamefont{M.}~\bibnamefont{Imada}},
  \bibinfo{journal}{Phys. Rev. B} \textbf{\bibinfo{volume}{71}},
  \bibinfo{pages}{045103} (\bibinfo{year}{2005}).

\bibitem[{\citenamefont{Miyake et~al.}(2009{\natexlab{a}})\citenamefont{Miyake,
  Aryasetiawan, and Imada}}]{MAI09}
\bibinfo{author}{\bibfnamefont{T.}~\bibnamefont{Miyake}},
  \bibinfo{author}{\bibfnamefont{F.}~\bibnamefont{Aryasetiawan}},
  \bibnamefont{and} \bibinfo{author}{\bibfnamefont{M.}~\bibnamefont{Imada}},
  \bibinfo{journal}{Phys. Rev. B} \textbf{\bibinfo{volume}{80}},
  \bibinfo{pages}{155134} (\bibinfo{year}{2009}{\natexlab{a}}).

\bibitem[{\citenamefont{Miyake and Aryasetiawan}(2008)}]{MiAr08}
\bibinfo{author}{\bibfnamefont{T.}~\bibnamefont{Miyake}} \bibnamefont{and}
  \bibinfo{author}{\bibfnamefont{F.}~\bibnamefont{Aryasetiawan}},
  \bibinfo{journal}{Phys. Rev. B} \textbf{\bibinfo{volume}{77}},
  \bibinfo{pages}{085122} (\bibinfo{year}{2008}).

\bibitem[{\citenamefont{Nakamura et~al.}(2008)\citenamefont{Nakamura, Arita,
  and Imada}}]{NAI08}
\bibinfo{author}{\bibfnamefont{K.}~\bibnamefont{Nakamura}},
  \bibinfo{author}{\bibfnamefont{R.}~\bibnamefont{Arita}}, \bibnamefont{and}
  \bibinfo{author}{\bibfnamefont{M.}~\bibnamefont{Imada}}, \bibinfo{journal}{J.
  Phys. Soc. Jpn.} \textbf{\bibinfo{volume}{77}}, \bibinfo{pages}{093711}
  (\bibinfo{year}{2008}).

\bibitem[{\citenamefont{Miyake et~al.}(2008)\citenamefont{Miyake, Pourovskii,
  Vildosola, Biermann, and Georges}}]{MPV+08}
\bibinfo{author}{\bibfnamefont{T.}~\bibnamefont{Miyake}},
  \bibinfo{author}{\bibfnamefont{L.}~\bibnamefont{Pourovskii}},
  \bibinfo{author}{\bibfnamefont{V.}~\bibnamefont{Vildosola}},
  \bibinfo{author}{\bibfnamefont{S.}~\bibnamefont{Biermann}}, \bibnamefont{and}
  \bibinfo{author}{\bibfnamefont{A.}~\bibnamefont{Georges}},
  \bibinfo{journal}{J. Phys. Soc. Jpn.} \textbf{\bibinfo{volume}{77}},
  \bibinfo{pages}{, Suppl. C,99} (\bibinfo{year}{2008}).

\bibitem[{\citenamefont{Miyake et~al.}(2009{\natexlab{b}})\citenamefont{Miyake,
  Aryasetiawan, and Imada}}]{MAI89}
\bibinfo{author}{\bibfnamefont{T.}~\bibnamefont{Miyake}},
  \bibinfo{author}{\bibfnamefont{F.}~\bibnamefont{Aryasetiawan}},
  \bibnamefont{and} \bibinfo{author}{\bibfnamefont{M.}~\bibnamefont{Imada}},
  \bibinfo{journal}{Phys. Rev. B} \textbf{\bibinfo{volume}{80}},
  \bibinfo{pages}{155134} (\bibinfo{year}{2009}{\natexlab{b}}).

\bibitem[{\citenamefont{Zhiqiang et~al.}(1991)\citenamefont{Zhiqiang, Helie,
  Wuyan, Zhi, and Qingqi}}]{ZHW+91}
\bibinfo{author}{\bibfnamefont{L.}~\bibnamefont{Zhiqiang}},
  \bibinfo{author}{\bibfnamefont{L.}~\bibnamefont{Helie}},
  \bibinfo{author}{\bibfnamefont{L.}~\bibnamefont{Wuyan}},
  \bibinfo{author}{\bibfnamefont{Z.}~\bibnamefont{Zhi}}, \bibnamefont{and}
  \bibinfo{author}{\bibfnamefont{Z.}~\bibnamefont{Qingqi}},
  \bibinfo{journal}{Solid State Commun.} \textbf{\bibinfo{volume}{79}},
  \bibinfo{pages}{791} (\bibinfo{year}{1991}).

\bibitem[{\citenamefont{Yang et~al.}(2011)\citenamefont{Yang, Yang, Chen, Ma,
  Han, Yang, Guo, Yan, Huang, Wu et~al.}}]{YYC+11}
\bibinfo{author}{\bibfnamefont{J.~B.} \bibnamefont{Yang}},
  \bibinfo{author}{\bibfnamefont{Y.~B.} \bibnamefont{Yang}},
  \bibinfo{author}{\bibfnamefont{X.~G.} \bibnamefont{Chen}},
  \bibinfo{author}{\bibfnamefont{X.~B.} \bibnamefont{Ma}},
  \bibinfo{author}{\bibfnamefont{J.~Z.} \bibnamefont{Han}},
  \bibinfo{author}{\bibfnamefont{Y.~C.} \bibnamefont{Yang}},
  \bibinfo{author}{\bibfnamefont{S.}~\bibnamefont{Guo}},
  \bibinfo{author}{\bibfnamefont{A.~R.} \bibnamefont{Yan}},
  \bibinfo{author}{\bibfnamefont{Q.~Z.} \bibnamefont{Huang}},
  \bibinfo{author}{\bibfnamefont{M.~M.} \bibnamefont{Wu}},
  \bibnamefont{et~al.}, \bibinfo{journal}{Appl. Physics Lett.}
  \textbf{\bibinfo{volume}{99}}, \bibinfo{pages}{082505}
  (\bibinfo{year}{2011}).

\bibitem[{\citenamefont{Koyama et~al.}(2008)\citenamefont{Koyama, Mitsui, and
  Watanabe}}]{KMW08}
\bibinfo{author}{\bibfnamefont{K.}~\bibnamefont{Koyama}},
  \bibinfo{author}{\bibfnamefont{Y.}~\bibnamefont{Mitsui}}, \bibnamefont{and}
  \bibinfo{author}{\bibfnamefont{K.}~\bibnamefont{Watanabe}},
  \bibinfo{journal}{Sci. Technol. Adv. Mater.} \textbf{\bibinfo{volume}{9}},
  \bibinfo{pages}{024204} (\bibinfo{year}{2008}).

\bibitem[{\citenamefont{Vast et~al.}(1992)\citenamefont{Vast, Siberchicot, and
  Zerah}}]{VSZ92}
\bibinfo{author}{\bibfnamefont{N.}~\bibnamefont{Vast}},
  \bibinfo{author}{\bibfnamefont{B.}~\bibnamefont{Siberchicot}},
  \bibnamefont{and} \bibinfo{author}{\bibfnamefont{P.~G.} \bibnamefont{Zerah}},
  \bibinfo{journal}{J. Phys.: Condens. Matter} \textbf{\bibinfo{volume}{4}},
  \bibinfo{pages}{10469} (\bibinfo{year}{1992}).

\bibitem[{\citenamefont{Yosida et~al.}(1965)\citenamefont{Yosida, Okiji, and
  Chikazumi}}]{YOC65}
\bibinfo{author}{\bibfnamefont{K.}~\bibnamefont{Yosida}},
  \bibinfo{author}{\bibfnamefont{A.}~\bibnamefont{Okiji}}, \bibnamefont{and}
  \bibinfo{author}{\bibfnamefont{S.}~\bibnamefont{Chikazumi}},
  \bibinfo{journal}{Prog. Theor. Phys.} \textbf{\bibinfo{volume}{33}},
  \bibinfo{pages}{559} (\bibinfo{year}{1965}).

\bibitem[{\citenamefont{Cinal et~al.}(1994)\citenamefont{Cinal, Edwards, and
  Mathon}}]{CEM94}
\bibinfo{author}{\bibfnamefont{M.}~\bibnamefont{Cinal}},
  \bibinfo{author}{\bibfnamefont{D.~M.} \bibnamefont{Edwards}},
  \bibnamefont{and} \bibinfo{author}{\bibfnamefont{J.}~\bibnamefont{Mathon}},
  \bibinfo{journal}{Phys. Rev. B} \textbf{\bibinfo{volume}{50}},
  \bibinfo{pages}{3754} (\bibinfo{year}{1994}).

\bibitem[{\citenamefont{Antropov et~al.}(2014)\citenamefont{Antropov, Ke, and
  Aberg}}]{AKA14}
\bibinfo{author}{\bibfnamefont{V.~P.} \bibnamefont{Antropov}},
  \bibinfo{author}{\bibfnamefont{L.}~\bibnamefont{Ke}}, \bibnamefont{and}
  \bibinfo{author}{\bibfnamefont{D.}~\bibnamefont{Aberg}},
  \bibinfo{journal}{Solid State Commun.} \textbf{\bibinfo{volume}{194}},
  \bibinfo{pages}{35} (\bibinfo{year}{2014}).

\bibitem[{\citenamefont{Ke et~al.}(2013)\citenamefont{Ke, Belashchenko, van
  Schilfgaarde, Kotani, and Antropov}}]{KBS13}
\bibinfo{author}{\bibfnamefont{L.}~\bibnamefont{Ke}},
  \bibinfo{author}{\bibfnamefont{K.~D.} \bibnamefont{Belashchenko}},
  \bibinfo{author}{\bibfnamefont{M.}~\bibnamefont{van Schilfgaarde}},
  \bibinfo{author}{\bibfnamefont{T.}~\bibnamefont{Kotani}}, \bibnamefont{and}
  \bibinfo{author}{\bibfnamefont{V.}~\bibnamefont{Antropov}},
  \bibinfo{journal}{Phys. Rev. B} \textbf{\bibinfo{volume}{88}},
  \bibinfo{pages}{024404} (\bibinfo{year}{2013}).

\bibitem[{\citenamefont{Montalti et~al.}(2006)\citenamefont{Montalti, Credi,
  Prodi, and Gandolfi}}]{book:MCP+06}
\bibinfo{author}{\bibfnamefont{M.}~\bibnamefont{Montalti}},
  \bibinfo{author}{\bibfnamefont{A.}~\bibnamefont{Credi}},
  \bibinfo{author}{\bibfnamefont{L.}~\bibnamefont{Prodi}}, \bibnamefont{and}
  \bibinfo{author}{\bibfnamefont{M.~T.} \bibnamefont{Gandolfi}}, in
  \emph{\bibinfo{booktitle}{Handbook of Photochemistry}}, edited by
  \bibinfo{editor}{\bibfnamefont{N.~J.} \bibnamefont{Turro}}
  (\bibinfo{publisher}{Taylor and Francis Group}, \bibinfo{address}{Boca Raton,
  FL}, \bibinfo{year}{2006}), p. \bibinfo{pages}{629}.

\bibitem[{\citenamefont{Zarkevich et~al.}(2013)\citenamefont{Zarkevich, Wang,
  and Johnson}}]{cm:ZWJ13}
\bibinfo{author}{\bibfnamefont{N.~A.} \bibnamefont{Zarkevich}},
  \bibinfo{author}{\bibfnamefont{L.-L.} \bibnamefont{Wang}}, \bibnamefont{and}
  \bibinfo{author}{\bibfnamefont{D.~D.} \bibnamefont{Johnson}},
  \bibinfo{journal}{preprint arXiv:1312.1988 [cond-mat.mtrl-sci]}
  (\bibinfo{year}{2013}).

\bibitem[{\citenamefont{Reim and Schoenes}(1990)}]{book:RS90}
\bibinfo{author}{\bibfnamefont{W.}~\bibnamefont{Reim}} \bibnamefont{and}
  \bibinfo{author}{\bibfnamefont{J.}~\bibnamefont{Schoenes}}, in
  \emph{\bibinfo{booktitle}{Ferromagnetic Materials}}, edited by
  \bibinfo{editor}{\bibfnamefont{E.~P.} \bibnamefont{Wohlfarth}}
  \bibnamefont{and} \bibinfo{editor}{\bibfnamefont{K.~H.~J.}
  \bibnamefont{Buschow}} (\bibinfo{publisher}{North-Holland},
  \bibinfo{address}{Amsterdam}, \bibinfo{year}{1990}),
  vol.~\bibinfo{volume}{5}, p. \bibinfo{pages}{133}.

\bibitem[{\citenamefont{Uba et~al.}(2000)\citenamefont{Uba, Uba, Antonov,
  Yaresko, Slezak, and Korecki}}]{UUA+00a}
\bibinfo{author}{\bibfnamefont{L.}~\bibnamefont{Uba}},
  \bibinfo{author}{\bibfnamefont{S.}~\bibnamefont{Uba}},
  \bibinfo{author}{\bibfnamefont{V.~N.} \bibnamefont{Antonov}},
  \bibinfo{author}{\bibfnamefont{A.~N.} \bibnamefont{Yaresko}},
  \bibinfo{author}{\bibfnamefont{T.}~\bibnamefont{Slezak}}, \bibnamefont{and}
  \bibinfo{author}{\bibfnamefont{J.}~\bibnamefont{Korecki}},
  \bibinfo{journal}{Phys. Rev. B} \textbf{\bibinfo{volume}{62}},
  \bibinfo{pages}{13731} (\bibinfo{year}{2000}).

\bibitem[{\citenamefont{Fuggle and Inglesfield}(1992)}]{book:FI92}
\bibinfo{author}{\bibfnamefont{J.~C.} \bibnamefont{Fuggle}} \bibnamefont{and}
  \bibinfo{author}{\bibfnamefont{J.~E.} \bibnamefont{Inglesfield}},
  \emph{\bibinfo{title}{Unoccupied Electronic States. Topics in Applied
  Physics}}, vol.~\bibinfo{volume}{69} (\bibinfo{publisher}{Springer},
  \bibinfo{address}{New York}, \bibinfo{year}{1992}).

\end{thebibliography}

\newcommand{\noopsort}[1]{} \newcommand{\printfirst}[2]{#1}
  \newcommand{\singleletter}[1]{#1} \newcommand{\switchargs}[2]{#2#1}

\end{document}